\documentclass{ws-rv961x669}
\usepackage{ws-rv-van}     
\usepackage{ws-rv-thm}     
\usepackage{subfigure}     
\makeindex
\usepackage{ulem}
\usepackage{float}

\newcommand{\Msun}{{\ensuremath{\mathrm{M}_\odot}}}

\begin{document}

\chapter[Black holes from stars]{Black holes as the end state of stellar evolution:\\ Theory and simulations}

\author[A. Heger, B. M\"{u}ller, I. Mandel]{Alexander Heger, Bernhard M\"{u}ller, \& Ilya Mandel}

\address{School of Physics and Astronomy, Monash University, Clayton, VIC 3800, Australia\\
alexander.heger@monash.edu, bernhard.mueller@monash.edu, ilya.mandel@monash.edu
}

\begin{abstract}
The collapse of massive stars is one of the most-studied paths to black hole formation. In this chapter, we review black
hole formation during the collapse of massive stars in the broader context of single and binary stellar evolution and the
theory of supernova explosions. We provide a concise overview of the evolutionary channels that may lead to black hole
formation -- the classical route of iron core collapse, collapse due to pair instability in very massive stars, and the
hypothetical scenario of supermassive star collapse. We then review the current understanding of the parameter space
for black hole formation and black hole birth properties that has emerged from theoretical and computational modelling
of supernova explosions and transient observations. Finally, we discuss what the intricate interplay between stellar evolution, stellar explosions, and binary interactions implies for the formation of stellar-mass black holes.
\end{abstract}


\body


\section{Introduction}

The concept of black holes has long been intimately tied to the theory of stellar evolution. While black hole spacetimes had been discovered and discussed as mathematical curiosities already shortly after the theory of general relativity was formulated, they were first seriously considered as astrophysical objects in the context of stellar collapse of sufficiently massive stars, most notably in the seminal work of Oppenheimer and Snyder\citep{oppenheimer_39}.
Stellar mass black holes formed as the end state of massive stars were also the first ones to be discovered (Cygnus X-1\citep{webster_72}).

Since the early days of general relativity and even since the days of Oppenheimer and Snyder, our understanding of stellar evolution and black hole formation has, however, changed considerably. Today, the black holes are well understood from the fundamental perspective of mathematical relativity, but the astrophysics of stellar-mass black hole formation still poses many deceptively simple questions despite considerable advances over the last few decades: Which massive stars form black hole, which ones form neutron stars? Do black holes form quietly, or are they sometimes born in supernova explosions and perhaps receive a ``kick'' in the process? How fast do newly born black holes spin? How do the fates of massive stars and the properties of black holes depend on their environment and how do they shape their environment through feedback processes?

In order to convey a comprehensive picture of our current theoretical understanding of black holes as the end states of stellar evolution, it is useful to approach the problem
from three different angles. In this Chapter, we first review the evolution of massive single stars up to the point of collapse in Section~\ref{sec:massive_stars}. We next discuss the current state of supernova theory with a focus on the outcomes of the collapse (neutron star formation, black hole formation with and without a supernova explosion) in Section~\ref{sec:SNBH}. The closely related topic of black hole birth properties is discussed in Section~\ref{sec:birth_properties}. Section~\ref{sec:binevol} reviews the interplay between black hole formation and binary evolution.  A brief summary can be found in Section~\ref{sec:summary}.

\section{Single star evolution up to the supernova}

\label{sec:massive_stars}

\begin{figure}[H]
    \centering
    \includegraphics[width=\textwidth]{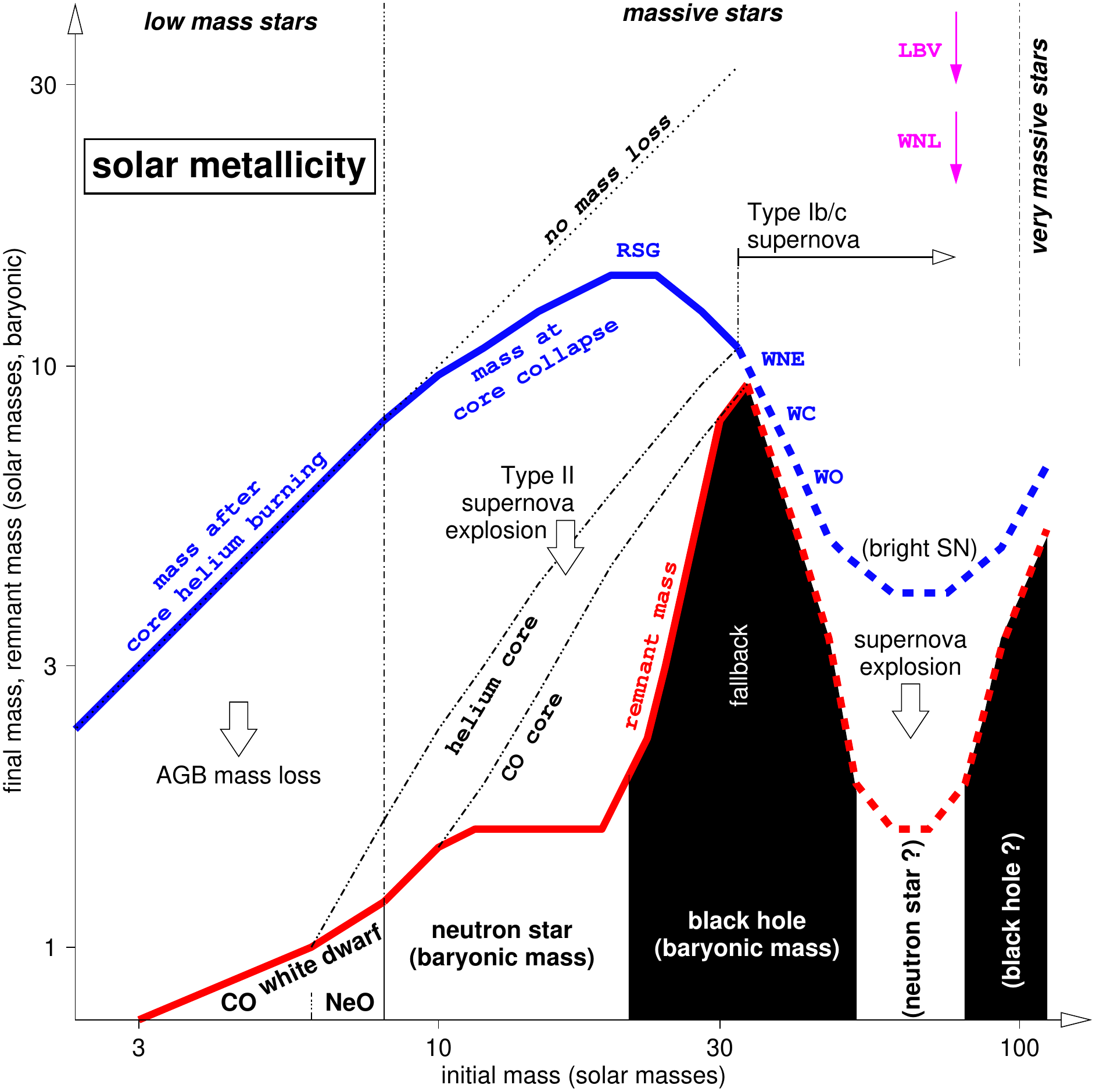}
    \caption{Cartoon sketch of remnant mass (\textsl{red line}) and stellar mass at ``time of remnant formation'' (\textsl{bluxe line}) as functions of initial mass for non-rotating single stars of about solar composition.  For stars of initial mass up to $\sim8\,M_\odot$ we show, crudely, the mass before the onset of the asymptotic giant branch (AGB) phase.  They leave behind carbon-oxygen (CO) or neon-magnesium-oxygen (NeO) white dwarfs.  \textsl{Dash-double-dotted} lines indicate the masses of the helium and CO cores at that evolution stage.  At an initial mass of about $30\,M_\odot$ the hydrogen envelope is lost due to stellar winds and the star becomes a Wolf-Rayet star (e.g., early-type Wolf-Rayet stars: WE, which could be nitrogen-rich sub-type WNE, carbon-rich sub-type WC, or oxygen-rich sub-type WO) prior to explosion - making Type I b/c supernovae, otherwise the star may explode as hydrogen-rich Type II supernova.   At higher masses the star may have strong mass loss already during the hydrogen-burning phase undergoing evolution as late-type nitrogen-rich Wolf-Rayet (WNL) star or a luminous blue variable (LBV).  At high initial mass the mass loss rates are highly uncerffig:tain and hence the final outcome is not reliably predicted, hence we use \textsl{dashed} lines.  Regimes of black hole formation may be interspersed with islands of neutron star formation even at relatively high masses.  We indicate one such island as a representation in this cartoon figure.
    \label{fig:MM1}
    }
\end{figure}

The life and fate of stars is predominately determined by their mass at birth (Figure~\ref{fig:MM1}).  Forming from a cloud of usually mostly molecular gas, objects with at least around $0.08$ times the mass of the sun ($\Msun$) can ignite hydrogen burning in their cores.  Stars with initial masses of $\lesssim1\,\Msun$ experience hydrogen burning powered by the proton-proton (PP) chains\cite{Iliadis07}, above that by the carbon-nitrogen-oxygen (CNO) cycle\cite{Iliadis07}.

The low-mass stars are faint red dwarf stars that can live a very long time: below a mass of $0.8\,\Msun$ they live as long as the current age of the universe; if not destroyed by some interaction, all single stars below this mass limit ever formed are still around.  Of these, stars with initial mass of up to $\sim0.6\,\Msun$ will end their lives as helium white dwarfs;  more massive stars can ignite helium burning, either in a core helium flash (up to $\sim2\,\Msun$ initial mass) or in less violent manner.  Stars with initial masses of up to $\sim6\,\Msun$ develop a degenerate carbon-oxygen (CO) core and leave behind a CO white dwarf (WD) of up to $1.1\,\Msun$\cite{KL14}.

More massive stars ignite carbon burning in their core.  Stars with initial masses of up to $\sim8\,\Msun$ cannot ignite further burning stages and leave behind oxygen-neon-magnesium (ONeMg) WDs of up to $\sim1.4\,\Msun$, just below the Chandrasekhar mass (Eq.~\ref{eq:MCh} with $Y_\mathrm{e}\sim0.5$ and $s\ll1\,k_\mathrm{B}/\mathrm{nucleon}$).  Just above this upper mass limit for WD formation the stellar evolution can become very complicated, e.g., leading to the formation of electron-capture supernovae\cite{Doherty17} or various off-centre advanced burning stages that may result in violent burning flashes, but usually lead to the formation of an iron core that undergoes core collapse\cite{WH15} (\S~\ref{sec:snmech}).  Above an initial mass of $\sim10\,\Msun$ the core becomes massive enough for stellar evolution to proceed in a more regular way (``textbook'' case of Figure~\ref{fig:KD}), also making an iron core that undergoes core collapse, leaving behind a neutron star or a black hole\cite{WHW02}.  We refer to such stars that make an iron core and collapse as \emph{massive stars}\cite{Heger12}.

At sufficiently low metallicity such that mass loss through winds can be neglected, we may expect the following mass regimes.
If the initial stellar mass exceeds $\sim90\,\Msun$, the stars encounter an instability in the equation of state after core carbon burning due to the production of electron-positron pairs, the \emph{pair instability}\cite{HW10}.  Stars above this mass limit we call \emph{very massive stars}\cite{Heger12}.  The pair instability cause thermonuclear powered pulses that expel the outer layers of the star\cite{Woosley17}.  We expect that these star usually should leave behind massive stellar black holes of up to $\sim45\,\Msun$.  For stars with initial masses above $\mathord\sim150\,\Msun$ the pulses can become so violent that the entire star is disrupted, usually during the first pulse, resulting in powerful supernova and no remnant\cite{HW02}.
Stars with initial masses above $\sim250\,\Msun-300\,\Msun$ encounter an instability due to photo-disintegration of heavy nuclei and helium during the pair-instability pulse, and collapse to a black hole rather than exploding\cite{HW02}.  The resulting back hole masses are expected to be at least $130\,\Msun$.  Observations of high-red shift quasars have lead to the speculation of the formation of primordial stars with the most extreme masses\cite{Woods17a}.  For such stars of primordial (i.e., Big Bang) composition, there is a hydrostatic upper mass limit of around $150\mathord,000\,\Msun$ due to a general relativistic instability\cite{Woods20,Haemmerle19,Haemmerle18}.  These stars would collapse to black holes on a thermal timescale, and stars beyond this mass limit we refer to as \emph{supermassive stars}.

\subsection{Massive star evolution}

\begin{figure}[H]
\includegraphics[width=\textwidth]{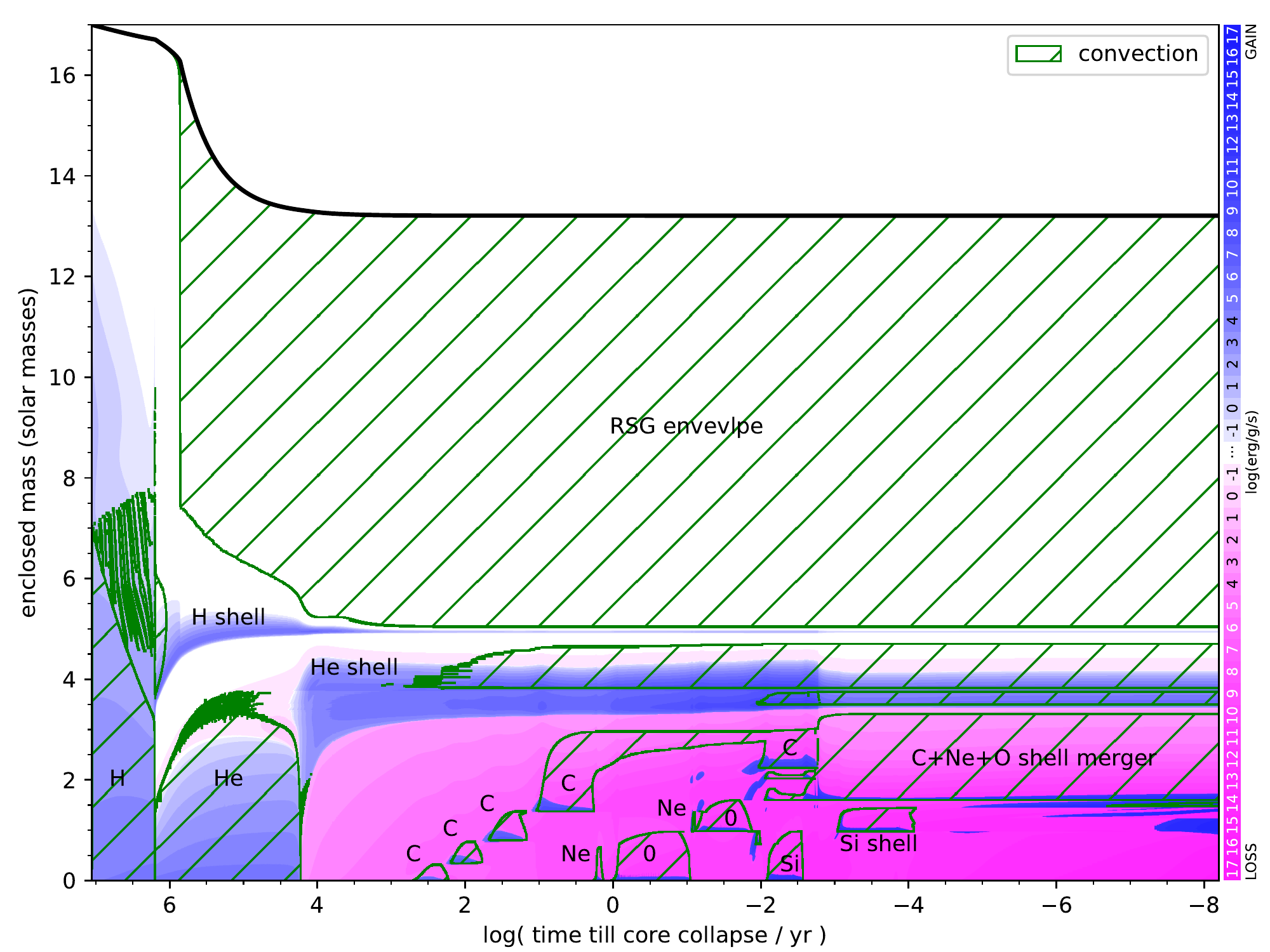}
\caption{Kippenhahn Diagram of the evolution of a non-rotating star of initially $17\,\Msun$ and solar initial composition.\cite{mueller_16a}  The $x$-axis shows the logarithm (base 10) of the remaining time in years until core collapse (core bounce) and the $y$-axis the enclosed mass (mass coordinate) for a given spherical shell.  The total mass of the star is indicated by a \textsl{solid black line} in the upper part of the figure; it is reduced by mass loss due to stellar winds.  Note that the mass loss rate is actually increasing toward the late evolution stages, however, the use of the logarithmic time scale on the $x$-axis stretches out the curve.  \textsl{Green hatching} indicates convective regions - energy is transported by convection, and at the same time keeps the connected regions close to chemically homogeneous.  At the end of the evolution the star becomes a red supergiant (RSG) and develops an extended convective envelope.  \textsl{Blue shading} indicates energy generation due to nuclear burning. \textsl{Purple shading} indicates net energy loss - energy carried away by neutrinos.  For both, energy production and loss, each level of shading corresponds to an increase of the specific energy generation or loss rate by one order of magnitude.  \textsl{Black labels} inside the figure indicate the respective burning phase, starting with hydrogen burning (\textsl{H}), followed by helium burning (\textsl{He}), then carbon (\textsl{C}), neon (\textsl{Ne}), oxygen (\textsl{O}) and finally (\textsl{Si}) burning.  When the fuel is depleted in the code, the burning can re-ignite in a shell.  We have marked the shells as such for hydrogen (\textsl{H shell}), helium (\textsl{He shell}), and silicon (\textsl{Si shell}); for carbon, neon and oxygen we just use the chemical symbol for both core and shell burning due to space constraints.  Note that the burning itself is usually highly concentrated toward the bottom of a convective burning shell.  After core silicon burning, this specific star experienced a merger of the second oxygen shell with the neon and carbon shells above (\textsl{C+Ne+O shell merger}).  This has significant impact on the structure of the star at the time of core collapse, with a fairly small core of $1.6\,\Msun$ surround by a high-entropy layer of low density, with the likely outcome being a neutron star not a black hole.  Earlier in the evolution, there is a sequence of thin convective regions - each framed by a solid line - above the hydrogen- and helium-burning cores.  These are a result of semiconvection, which may cause only a modest amount of mixing.\label{fig:KD}}
\end{figure}

From its formation, the evolution of a massive star can be sketched as a continuing path toward increasingly higher central temperature and density, while central entropy decreases.  This overall path, however, is interrupted by stages in thermal equilibrium, powered by nuclear burning.  This nuclear burning, in turn, can trigger local instability and drive convection.  In the advanced burning stages, when the temperature exceeds some $10^9\,\mathrm{K}$, energy loss due to thermal neutrinos can drive the evolution timescale.  Figure~\ref{fig:KD} shows the evolution of these quantities as a function of time and of location inside the star.  The different burning stages in a massive star, in sequence, are:

\begin{itemize}
\item \textbf{Hydrogen burning.}  In massive stars this is powered by the CNO cycle, $^{12}$C(p,$\gamma$)$^{13}$N($\beta^+$)$^{13}$C(p,$\gamma$)$^{14}$N(p,$\gamma$)$^{15}$O($\beta^+$)$^{15}$N(p,$\alpha$)$^{12}$C and converts hydrogen to helium, releasing about $6.5\,\mathrm{MeV}$ per nucleon in nuclear binding energy.\footnote{We use a notation common in nuclear astrophysics to show the reaction.  Heavy ions are usually shown outside the bracket.  Within the bracket there is either one group or two groups separated by a comma.  If there is only one group, it either shows what is emitted due to a spontaneous process or the process itself.  For example, $\beta^+$ decay emits a positron and an electron neutrino, which we also could have written as $e^+\nu_\mathrm{e}$.   If there are two groups separated by a comma, the first group is the ``ingoing'' channel, i.e., the reactant(s), and the second group is the ``outgoing'' channel, i.e., products.  In case there is only an ingoing channel, it is customary to write $\gamma$ for the outgoing channel.}  About $7\,\%$ of that energy is radiated away in the form of neutrinos.  This energy generation is about one order of magnitude larger per unit mass than any of the later nuclear burning stages, while the star typically also is less luminous, making core hydrogen burning, often referred to as the \textsl{main sequence}, the longest evolutionary phase of massive stars.  When hydrogen is exhausted in the core, it usually continues to burn in a shell outside the helium core.  In the advanced burning stages beyond helium burning (see below), the hydrogen shell may become inactive, it may be ``dredged up'' (mixed with) a convective envelope at the low-mass end for core collapse supernovae, of be lost due to stellar winds or eruptions in the most massive stars.

\item \textbf{Helium burning.}  This phase is started by the ``triple alpha'' reaction, $3\,{}^4$He$\;\rightarrow{}^{12}$C.  As $^{12}$C accumulates and $^4$He is depleted, the $^{12}$C($\alpha$,$\gamma$)$^{16}$O reaction starts to dominate.  Since both reactions have about the same temperature dependence at typical helium burning conditions but different dependence on density, the resulting final carbon-to-oxygen ratio varies.  Low mass stars have higher densities and produce a larger carbon mass fraction.   The outcome also depends sensitively on the uncertainty in these two reaction rates, which still have relevant experimental uncertainties.  At the very end of core helium burning, some trace of $^{20}$Ne or heavier alpha nuclei (nuclei that are multiples of alpha particles) may be made.  After helium burning is depleted in the core, it can re-ignite in a shell outside the CO core.  For sufficiently large initial stellar masses, that shell can become convective and even entrain some hydrogen from the envelope.  At the low-mass end for supernovae, instead, the shell may be dredged up by the envelope (see above).  As helium burning releases only about $10\,\%$ of the energy per nucleon compared to hydrogen burning, the core helium burning phase typically lasts about only $10\,\%$ of the time of core hydrogen burning.  There is some contribution of shell hydrogen burning during that phase, but the stars also are typically more luminous.

\item \textbf{Carbon burning.}  Carbon burning predominately starts by the $^{12}\mathrm{C}+{}^{12}\mathrm{C}$ reactions.  The resulting compound nucleus of $^{24*}$Mg
excited state de-excites by particle emission of neutrons, protons, and alpha particles.\footnote{The asterisk indicates an excited state of the $^{24}$Mg nucleus.}  These make secondary reactions.  The typical outcome is production of $^{20}$Ne and $^{16}$O.  Most importantly, whether core carbon burning proceeds as a convective phase, as in Figure~\ref{fig:KD}, or radiative depends on the carbon mass fraction left behind by core helium burning.  When the burning is convective, there is more time for both loss of entropy due to neutrinos as well as for weak decays leading to a lower proton-to-neutron ratio (i.e., lower $Y_\mathrm{e}$, see below).  This leads to a transition in the pre-supernova structure, with typically stars on the high-mass side of this transition being less likely to explode.  At the transition itself, we find a sequence of many small shells, leading to many discontinuous changes in the stellar structure around the transition mass, which is about $18\,\Msun$ for the solar composition models of Reference~\refcite{mueller_16a}.  During core carbon burning, the star emits about $10\mathord,000$ times more energy in neutrinos than in visible light.  From this point on, the star has effectively become a \emph{neutrino star}.   At the low-mass end of the core collapse mass regime, carbon may ignite off-centre and burn inward in a convectively-bounded flame. \cite{WH15}

\item \textbf{Neon burning.}  This phase is powered by a pair of reactions, $^{20}$Ne($\gamma$,$\alpha$)$^{16}$O and $^{20}$Ne($\alpha$,$\gamma$)$^{24}$Mg, effectively burning $^{20}$Ne to $^{24}$Mg and $^{16}$O.  The first of the two reactions is endothermic, but the second reaction makes up for that by releasing about twice as much energy as is needed to trigger the first reaction. It is usually a rather brief and ``flashy'' phase, being induced by a photo-disintegration reaction that causes strong positive self-feedback.  It occurs briefly before each of the oxygen burning phases, and often burns with the convective region less extended in mass than the later oxygen burning.   Similarly to carbon burning, neon and subsequently oxygen may ignite off-centre and burn inward in a convectively-bounded flame at the low-mass end of the core collapse mass regime.\cite{WH15}

\item \textbf{Oxygen burning.}  The nuclear reactions powering oxygen burning proceed very similar to carbon burning.  The $^{16}\mathrm{O}+{}^{16}\mathrm{O}$ reaction produces a compound nucleus of $^{32*}$S that predominately de-excites by particle emission of neutrons, protons, and alpha particles that then induce secondary reactions.  The outcome is a mixture overwhelmingly consisting of $^{28}$Si and $^{32}$S.  The small mass fraction $^{24}$Mg, usually around $10\,\%$, left by neon burning is consumed at the beginning of oxygen burning by photo-disintegration reactions.  With a typical oxygen mass fraction of around $80\,\%$, the phase is relatively powerful and extended compared to neon burning.  There are usually one or two oxygen burning shells prior to core collapse, and during collapse oxygen burning can become ``explosive'' with burning timescales of a fraction of a second.  It typically sets a specific entropy of $\gtrsim4\,k_\mathrm{B}/\mathrm{nucleon}$, associated with a jump in density that can have a critical role in inducing the neutrino-powered core collapse supernova mechanism in some mass ranges.

\item \textbf{Silicon burning.} Silicon burning is dominated by a sequence of photo-disintegrations and $\alpha$ captures.  It usually lasts for just days.  During the burning, electron captures convert protons (inside nuclei) into neutrons, decreasing the electron fraction,\footnote{This equals the fraction of protons relative to all nucleons (neutrons and protons).} $Y_\mathrm{e}$, below $0.5$, and leaving behind a mixture of iron group isotopes.  For not too massive stars, the core silicon burning usually comprises a convective core of about $1.05\,\Msun$, followed by at least one silicon burning shell.  At the low-mass end, below $\sim12\,\Msun$ initial mass, more complicated burning sequences may occur, e.g., silicon shell burning igniting not at the bottom of silicon shell, leaving behind a layer of unburnt silicon between two layers of iron.

\item \textbf{Iron core collapse.}  The silicon continues to burn in shells until the critical mass for collapse (Eq.~\ref{eq:MCh}) is exceeded.  At this stage the iron core is very hot and in nuclear statistical equilibrium (NSE), i.e., nuclear reactions are very fast compared to the evolution time-scale of the star. Then further electron captures combined with photo-disintegration soften the equation of state and lead to the collapse of the iron core at about a quarter of the free-fall acceleration.  From the time that an infall velocity of $1\mathord,000\,$km/s is reached, core bounce typically ensues within a fraction of a second.

During the collapse, remaining layers of silicon as well as the bottom of the oxygen layer may undergo very fast, ``explosive'' burning on a timescale shorter than the collapse timescale or any convective or sonic timescale.  This may seed large asymmetries in the infall flow.
\end{itemize}

Nuclear burning predominately proceeds in distinct convective layers, which has a pronounced effect on distribution of outcomes as a function of initial mass.  For example, if at the end of silicon burning the iron core mass was just below the critical mass for collapse, another shell of convective silicon burning would be required, and, due to its finite mass, would lead to a much larger iron core, well above the critical mass.
A discontinuous jump occurs.  Similarly, earlier burning stages and their shells require a minimum mass for ignition, leading to many jumps in properties of the pre-supernova structure and the resulting outcomes (Figure~\ref{fig:landscape}).  In particular, the transition of core carbon burning from \textsl{(i)} being convective in large shells, to \textsl{(ii)} a sequence of very small convective shells, to \textsl{(iii)} radiative burning at $\sim18\,\Msun$ initial mass, and the non-linear impact of these shells on the subsequent more advanced core and shell burning stages, leads to many changes within a small range of initial mass, resembling almost a ``chaotic'' behaviour\footnote{although not ``chaotic'' in the mathematical sense}.
Stellar evolution solves a system of tightly-coupled non-linear partial differential equations with non-trivial inhomogeneous functions, and hence a first-principle prediction of outcomes such as ``compactness'' at the presupernova stage is difficult.  That said, the predictions between different codes do usually agree well given the same input physics\cite{Jones15, SW14}.  \label{par:C_shell}

\begin{figure}[H]
    \centering
    \includegraphics[width=\textwidth]{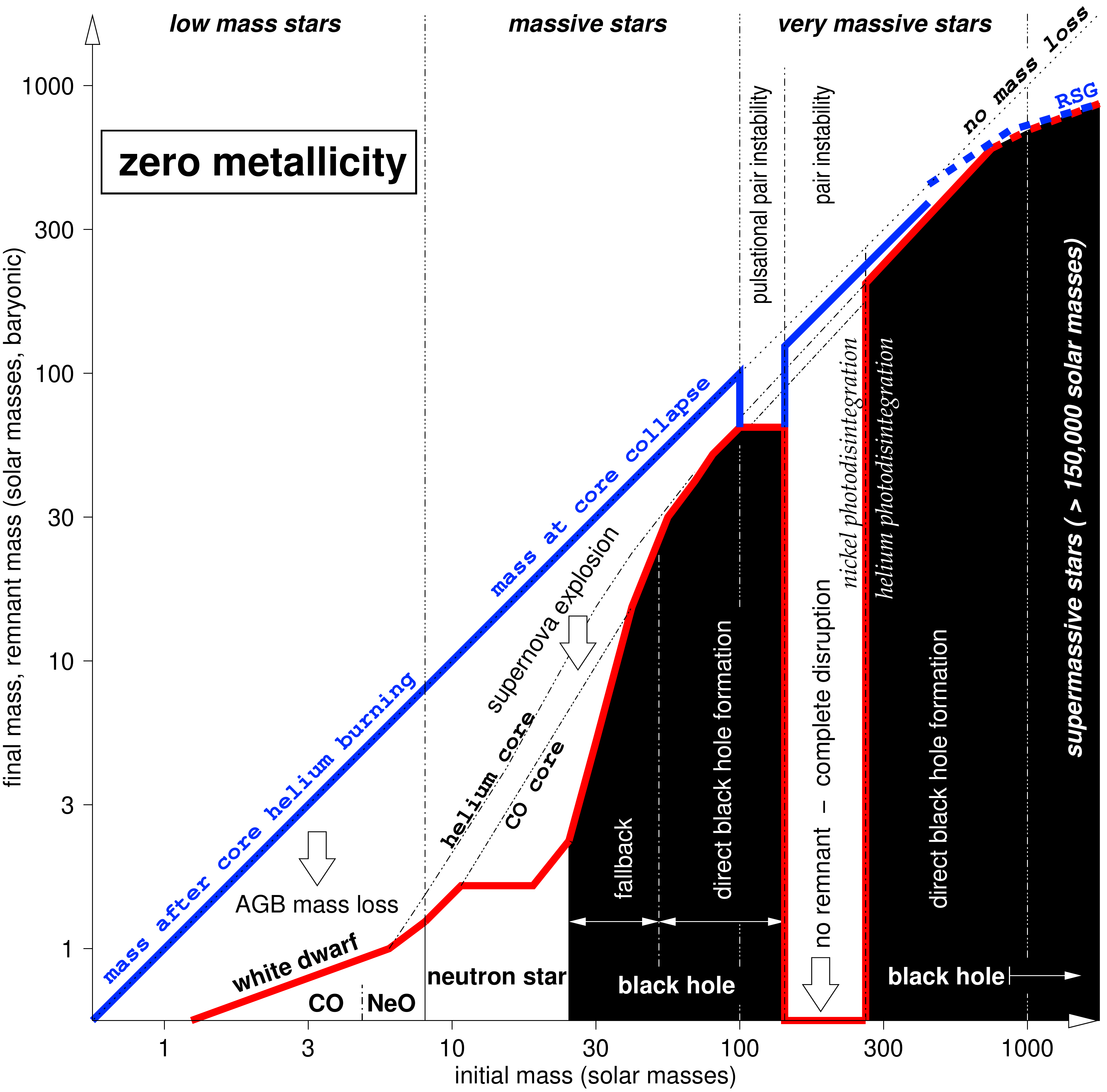}
    \caption{
    Similar to Figure~\ref{fig:MM1} but for Population III stars - stars made of pristine primordial material as left behind directly by the Big Bang.  In contrast to solar-composition stars, Population III stars may lose less mass due to stellar winds.  They reach large core masses at the end of their evolution and can encounter the pulsational pair instability for initial masses between $90\,M_\odot$ and $140\,M_\odot$ solar masses that sheds the outer layers down to some threshold (Figure~\ref{fig:ppsn}), and full pair instability between $140\,M_\odot$ and $260\,M_\odot$ that may not leave behind any remnant at all.  For higher masses, the star collapses directly to a large black hole.  Above an initial mass of some $150\mathord,000\,M_\odot$ (not shown) there does not exist a long-lived solution of a star in thermal equilibrium.   These  ``supermassive'' stars would collapse into supermassive black holes that could be the ideal seeds for quasars at high redshift.
\label{fig:MM3}
    }
\end{figure}

\begin{figure}
    \centering
    \includegraphics[width=\textwidth]{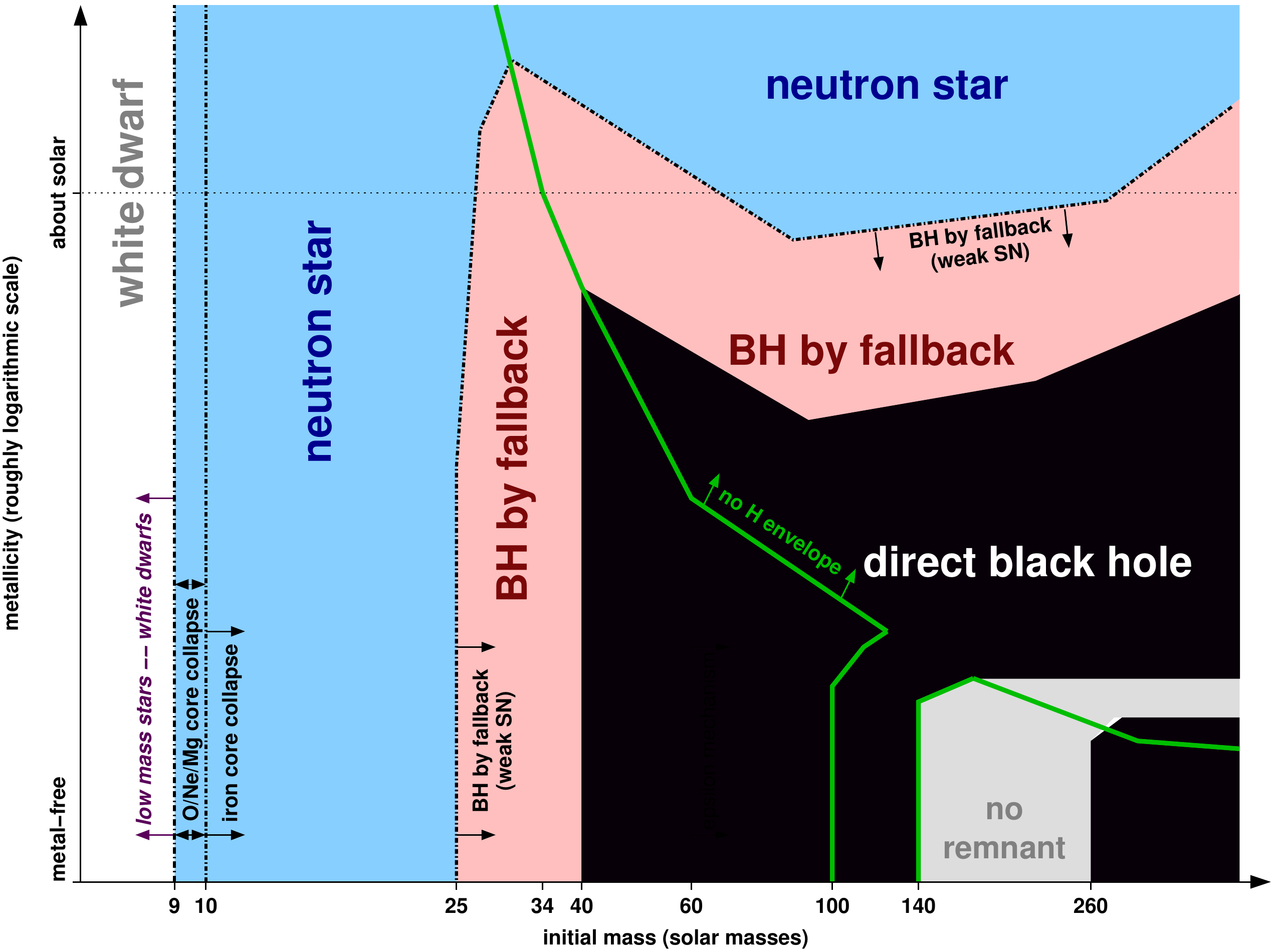}
    \caption{
    Schematic for regimes of compact remnants of single stars as a function of initial mass ($x$-\textsl{axis}) and metallicity ($y$-axis). The metallicity scale is approximate due largely to the uncertainty in stellar mass loss.  We show white dwarfs (\textsl{white shading}), neutron stars (\textsl{light blue shading}), black holes due to fallback after an initial explosions (\textsl{light red shading}), direct collapse (\textsl{black shading}), and a regime of no remnants (\textsl{light gray shading}) due to pair instability supernovae.  The green line shows where mass loss may remove the hydrogen-rich envelope prior to explosion of the star; for low metallicity between $100\,M_\odot$ and $140\,M_\odot$ initial mass it is due to pulsational pair instability supernovae, otherwise due to stellar winds.
    \label{fig:MZ}
    }
\end{figure}

\subsection{Stellar evolution physical parameters and uncertainties}

In this section we briefly discuss some key dependencies on input physics for single massive star evolution and their uncertainties.  Each of these topics could fill books by themselves.  Uncertainties due to binary evolution are discussed in Section~\ref{sec:binevol}.

\subsubsection{Metallicity and Mass loss}

Next to the initial mass of the star, metallicity is one of the key factors impacting stellar evolution, stellar populations and outcomes.  For example, the first generation of stars forming from metal-free pristine gas as left over by the Big Bang, so-called Population III stars, would not have strong molecular cooling from complex molecules, resulting in higher Jeans mass and hence typically making more massive stars than what we find in the more metal-rich present-day universe\cite{larson_98,abel_00,bromm_04}.  Whereas the actual typical initial mass function, which may depend on environment, is still subject to much debate, some constraints have already become apparent: for example, we have not found any Population III stars to date, whereas all single stars with initial masses of $0.8\,M_\odot$ or below should still be around today as their lifetime would exceed the current age of the universe.  This could be taken as an indication that the first generation of stars was typically more massive than stars today, although this fact provides no actual constraints on the more massive stars and black hole production.

Beyond the potentially different initial mass function of Population III stars, their evolution is also different.  Massive Population III stars, which burn hydrogen via the CNO cycle, have to first produce carbon by the triple alpha reaction in a primary fashion.  They contract to high enough densities and temperatures to start the helium fusion, until a carbon mass fraction of $\lesssim10^{-9}$ is reached.    Then they produce enough energy by hydrogen burning to balance energy losses from the surface and halt contraction.  The lack of initial metals, when preserved in the surface layers, likely leads to much reduced mass loss due to stellar winds.  The stars may retain most of their initial mass and may follow different evolutionary paths to stars of solar initial composition, for example, producing pair instability supernovae (see below) or more massive black holes.   Figure~\ref{fig:MM3} depicts a schematic for the potential evolution of non-rotating single Population III stars\cite{heger_02}.

Other than some peculiarities in the burning of the metal-free and very low metallicity stars, mass loss is one of the key drivers for different evolution pathways and outcomes for massive stars: it reduces the mass of the star, shrinking the resulting core sizes, or even ejects the entire hydrogen envelope, changing the observational astronomical supernova type as well as remnant type and mass (Figure~\ref{fig:MZ})\cite{heger_03}.

The mass loss itself, however, is still highly uncertain.  Significant efforts exist to constrain them from theory and observation.    Asymmetries and clumping in winds, dust in red supergiants\cite{humphresy_20,humpphreys_22}, line driving in hot stars and optically thick winds\cite{castor_75,abbott_80,friend_83,gayley_95,kudritzki_00,smith_14,graefner_17,Vink_22} with their metallicity dependence\cite{kudritzki_02,Vink_05}, episodic mass loss such as luminous blue variables (LBV) or giant eruptions such as observed in Eta Carina (Figure~\ref{fig:MM1})\cite{smith_06,humphreys_16,humphreys_17} constitute sizeable challenges in these efforts.

\subsubsection{Mixing, rotation, and magnetic fields}

Mixing and transport processes play a key role in the evolution of stars.  Foremost among these is the transport of energy due to radiation, conduction, or advection by fluid flows such as convection when other processes (radiation, conduction, mechanical work, and neutrino losses) are insufficient. Most critically, fluid flows also lead to transport and mixing of different layers in the star, bringing fuel into burning regions and the products of nuclear burning to the surface of the star.  Whereas convection is reasonably well described by the mixing length theory\cite{boehm_58}, many questions remain with regards to double diffusive instabilities such as semiconvection\cite{langer_83,spruit_92,geraud_18} and thermohaline convection\cite{stern_60,ulrich_72,kippenhahn_80,denissenkov_10}.  Semiconvection and thermohaline convection occur in dynamically stable regions, with no Rayleigh-Taylor instability, but with buoyancy due to composition gradients and thermal gradients pointing in opposite directions, with the stabilizing gradient dominating.  They are called ``double diffusive'' instabilities because the diffusion coefficients for ``heat'' and ``composition'' are vastly different in typical stellar conditions, with heat diffusing much faster than chemical species (atomic nuclei).  In particular, in semiconvection, a destabilizing temperature gradient is stabilised against dynamic instability by a larger stabilising composition gradient.  Secular instability leads to layer formation and eventually mixing driven by a slow exchange though the layer boundaries and through the merging of the layers.  In thermohaline convection, a destabilising composition gradient (``heavy'' material above ``lighter'' material) is stabilised against dynamical instability by a larger stabilising temperature gradient.  ``Fingers'' of larger chemical buoyancy may form and cool as they sink, though coherent structures, may be destroyed by turbulence and (differential) rotation\cite{denissenkov_10}.  These mixing and transport processes, however, are inherently three-dimensional with a vast range of scales, requiring theoretical insight to model them in lower dimensions over the entire evolution of a star.

A further quantity that is transported in stars is angular momentum.  Rotation can be an important aspect of stellar evolution\cite{maeder_00}.  In particular, massive stars may be spinning rapidly throughout much of their lives.  In close and interacting binary stars, stellar and (vast) orbital angular momentum can be exchanged, which can spin stars up or down.  For single stars, mass loss due to stellar winds can lead to significant braking as the surface layer of the star has the highest specific moment of inertia.  For magnetic stars -- usually stars with convective envelopes such as the Sun -- magnetic fields force escaping wind particles to remain in co-rotation with the surface out to large distances from the star, which results in particularly efficient loss of angular momentum.  This process is known as magnetic braking\cite{weber_67}.  Magnetic fields in the stellar interior can also have significant impact on angular momentum transport\cite{spruit_02} and the resulting final spin of the stellar core at the time of core collapse\cite{heger_05,woosley_06a}.

Stellar rotation deforms the stars, leading to different temperature gradients from the core to the surface at the pole than at the equator.  This can drive fluid flows, so-called meridional circulation due to its axisymmetry\cite{zahn_92}.  For very rapid rotation, the mixing can be faster than the nuclear burning, leading  to chemically homogeneous evolution\cite{HLW00,yoon_12}.  When the star reaches the end of core hydrogen burning, also outer layers are also depleted in hydrogen, altering stellar structure, mass, and angular momentum loss.

\subsubsection{Nuclear physics uncertainties}

Stellar structure and evolution are driven by nuclear physics.  Nuclear physics defines the different evolution stages.  Nuclear structure  -- and hence nuclear reactions -- are very complicated strongly-interacting quantum many body systems and hence accurate first-principle calculations of stellar structure is quite challenging.  On the other hand, nuclear physics experiments \emph{in the relevant -- usually low-energy -- regime} to directly measure reactions are also very challenging.  They require very low background experimental environments.  This becomes clear when you consider that stars may take millions of years for some burning phases, hence little happens during a human lifetime. The uncertainties can become an issue when one reaches branching points in the nuclear reaction flows or has competing processes, and this can have significant impact on stellar nucleosynthesis.

Nuclear reaction rates have very high temperature sensitivity, e.g., $\sim T^{40}$ for helium burning at typical hydrostatic helium burning temperatures.  This means that if, e.g., a rate was changed by a factor two, changing temperature by a factor $2^{-1/40}$ or $1.75\%$ would result in the same burning rate.  For helium burning in particular, however, there are two competing reactions, triple alpha and $^{12}$C($\alpha$,$\gamma$)$^{16}$O, that determine the carbon mass fraction at the end of core helium burning, which, in turn, impacts the carbon burning phases and ultimately the final stellar fate.  Since both reactions have about the same temperature dependence at the relevant temperatures, it is the difference in their density dependence that also plays a role.  Realistically, we would like to know these two rates to within some 5\% accuracy\cite{tur_07,west_13} but measurements are hard\cite{kibedi_20}.

Whereas carbon production is the most prominent and likely most impactful\cite{takahashi_18,farmer_20,woosley_21}, other reactions also have their key roles.  This includes light reactions in the CNO cycles, branching points in carbon burning\cite{tang_15,sukhbold_20}, and weak reaction rates in silicon burning and in the iron core\cite{heger_01,langanke_21}.

\section{Stellar collapse leading to the formation of black holes}

\label{sec:SNBH}

\subsection{The core collapse supernova mechanism}
\label{sec:snmech}
For massive stars with helium core masses below the somewhat uncertain threshold value for
pair instability supernovae (Section~\ref{sec:pisn}),
compact object formation proceeds through core collapse after hydrostatic burning stages
up to the formation of an iron core. In addition, there may be a narrow channel of less massive supernova progenitors that proceed through carbon burning to form a degenerate O-Ne-Mg core and already undergo
dynamical collapse at this stage due to electron captures
on ${}^{20}\mathrm{Ne}$ and ${}^{24}\mathrm{Mg}$ \citep{nomoto_84,nomoto_87,jones_14,nomoto_17}.
This progenitor channel invariably produces neutron stars, however. For low-mass
non-rotating progenitors, baryonic remnant masses are expected to be close to $1.37 M_\odot$ (resulting in a gravitational mass of about $1.25 M_\odot$). In the special case of accretion-induced collapse of rotating white dwarfs,
neutron stars masses may be higher \citep{yoon_05}. As far as black-hole formation is concerned,
we need only consider the standard scenario of iron core collapse.

In progenitors with an iron core, collapse occurs once the degenerate core
reaches its effective Chandrasekhar mass,
$M_\mathrm{Ch}$, which is given by\cite{timmes_96}
\begin{equation}
M_\mathrm{Ch}=
1.45 M_\odot \left(\frac{Y_\mathrm{e}}{0.5}\right)^2 \left[1+\left(\frac{s}{\pi Y_\mathrm{e} k_\mathrm{B}/\mathrm{nucleon}}\right)^2\right]
\label{eq:MCh}
\end{equation}
including finite-temperature corrections.  Here $Y_\mathrm{e}$ is the electron
fraction, and $s$ is the specific entropy $s$ of the core. While
$Y_\mathrm{e}\approx 0.44$  does not vary strongly across
progenitors, variations in core entropy between $0.5\,k_\mathrm{B}/\mathrm{nucleon}$ for the lowest-mass stars and $1.5\,k_\mathrm{B}/\mathrm{nucleon}$ at high masses (close to the pulsational pair instability regime) lead to
substantial variations in the final iron core mass. The contraction of the
core accelerates into a runaway collapse on a free-fall timescale because
electron captures on heavy nuclei and the small number of free protons further
reduce the degeneracy pressure; in case of higher core entropy, the reduction
of radiation pressure by photodisintegration of heavy nuclei is also relevant.

At core densities of about $10^{12}\, \mathrm{g}\, \mathrm{cm}^{-3}$, neutrinos become trapped
and electron captures can no longer reduce the lepton number of
the core. At this stage, the electron fraction of the core has decreased
to $Y_\mathrm{e}\approx 0.25$ \cite{langanke_03}. Due to the loss of
lepton number, the effective Chandrasekhar mass of the core shrinks
during collapse. Only the inner core maintains sonic contact and
remains in homologous collapse until it reaches and overshoots
nuclear saturation density \citep{goldreich_80}. Due to the stiffening of the
equation of state above nuclear density, the inner core rebounds (``bounce''),
and a shock wave is launched as the rebounding inner core crashes
into the supersonically collapsing shell of the outer core. At
core bounce, the newly formed compact remnant is still small with
a mass of around $0.45 M_\odot$ (somewhat dependent on the nuclear equation of state). In modern supernova simulations
using up-to-date stellar progenitor models, the iron core collapse of massive
stars therefore never results in prompt black hole formation;
there is always at least a transient proto-neutron star phase.

\begin{figure}
    \centering
    \includegraphics[width=0.75\linewidth]{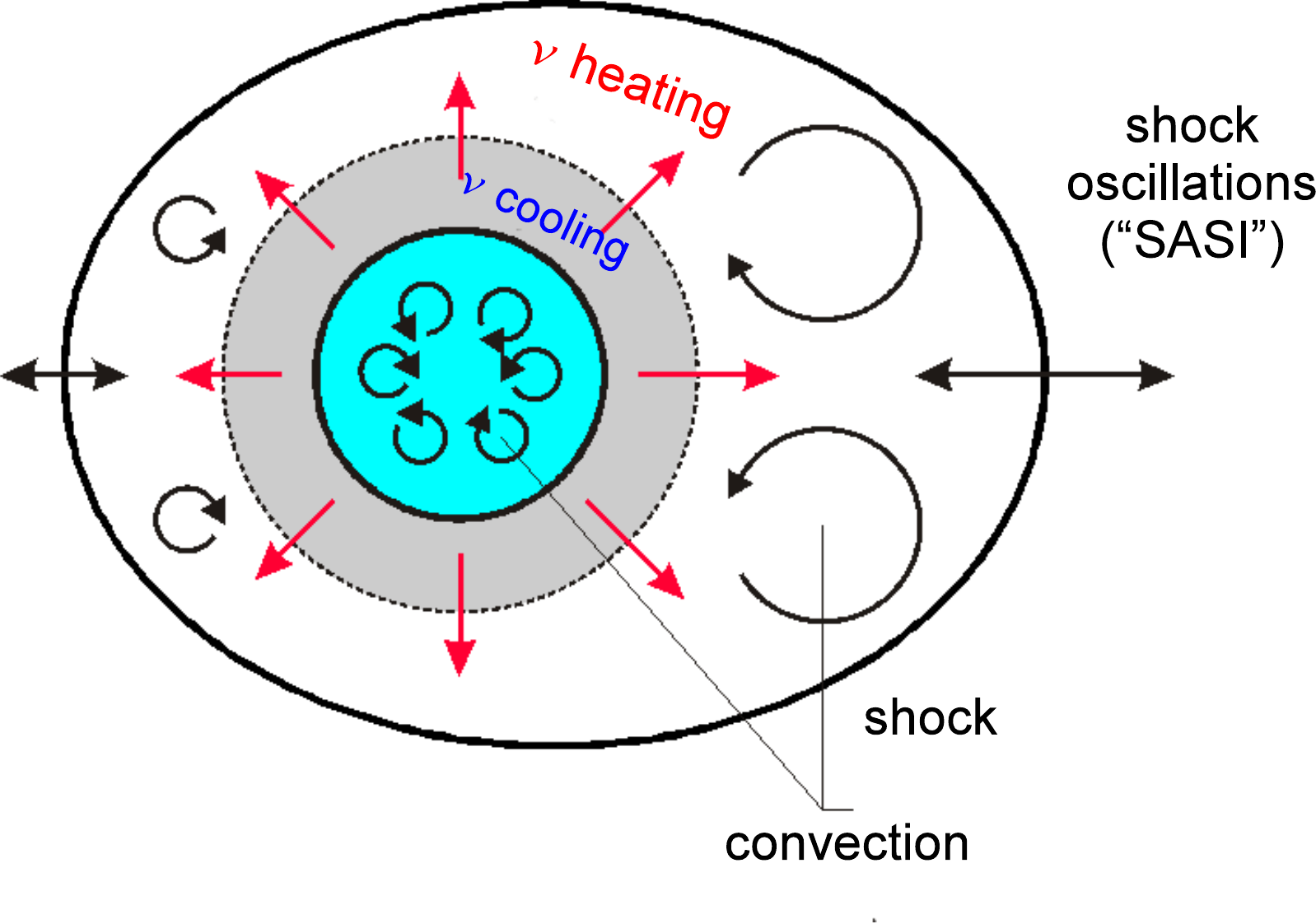}
    \caption{Sketch of the supernova core prior to shock revival in a neutrino-driven supernova. A fraction of the neutrinos emitted from the proto-neutron star (grey, cyan) are reabsorbed in the ``gain region'' behind the stalled shock further out. Neutrino heating drives convection in the gain region, and in addition, the standing accretion shock instability (SASI) can lead to large-scale shock oscillations. Neutrino heating in conjunction with the violent non-spherical fluid motions eventually drives runaway shock expansion. Neutrino cooling also drives convection inside the proto-neutron star.
    Figure from M\"{u}ller (2017)\cite{mueller_17b}
    reproduced with permission, \textcopyright{Cambridge University Press}.
    }
    \label{fig:neutrino}
\end{figure}

As the shock propagates through the outer core, its energy is quickly
drained by dissociation of heavy nuclei in the infalling shells
and by neutrino losses \cite{mazurek_82,burrows_85,bethe_90}. Within milliseconds after bounce,
the shock turns into an accretion shock
(i.e., the post-shock velocity becomes negative) that still reaches a radius
of $100\texttt{-}200\,\mathrm{km}$ and then retracts again. Unless
the shock is ``revived'', ongoing accretion will eventually lead
to black hole formation (although part of the star may still be ejected
in this case; see Section~\ref{sec:mass_loss}).

Various mechanisms for shock revival have been explored in the literature.
In the neutrino-driven paradigm (Figure~\ref{fig:neutrino}), the partial
absorption of neutrinos from the proto-neutron star behind the shock
increases the post-shock pressure to allow the shock to expand. Once the
volume and mass of the heating region have been increased sufficiently,
a runaway feedback cycle of stronger heating and shock expansion can occur \cite{thompson_00,janka_01,buras_06b}.
Since  most of the neutrino emission feeds on the accretion power of
the infalling material, the neutrino-driven mechanism is somewhat self-regulating
and will roughly pump energy into the post-shock matter until the incipient
explosion is sufficiently energetic to terminate further accretion \citep{mueller_16a}.
Except for the least massive supernova progenitors, hydrodynamic instabilities
such as buoyancy-driven convection \citep{herant_94,burrows_95} and shock oscillations \citep{blondin_03}
(standing accretion shock instability, SASI) play a crucial role in supporting
neutrino heating by providing additional turbulent pressure \citep{burrows_95,murphy_12} and
transporting hot material from the proto-neutron star to the shock \citep{janka_96}.

In the case of rapid progenitor rotation, an explosion may instead be
driven by magnetic fields that tap the rotational energy of the
proto-neutron star\cite{akiyama_03,burrows_07,winteler_12,moesta_14b}(magnetorotational mechanism). Similar
mechanisms may also operate after black hole formation in collapsar
disks (Section~\ref{sec:hne}), but it is convenient to distinguish
the collapsar scenario from magnetorotational explosions proper. The interplay
of rotation and magnetic fields in the progenitor and after collapse
are still topics of ongoing research, and hence it is far from clear
when the magnetorotational mechanism can operate. Nevertheless, some
robust features of the magnetorotational mechanism can be identified. The
energy reservoir for magnetorotational explosions is determined by
the free rotational energy in \emph{differential} rotation
of the proto-neutron star on short timescales \citep{burrows_07} and by its entire rotational energy on
longer timescales \citep{akiyama_03}. Simulations of successful magnetorotational
explosions tend to exhibit bipolar jet-like outflows \cite{burrows_07,winteler_12,moesta_14b} that
suggest an association with broad-lined Ic supernovae whose polarisation indicates
a nearly axisymmetric bipolar structure of the bulk of the ejecta\cite{woosley_06,wang_08}.

In recent years, phase-transition driven explosions have been considered
as yet another alternative explosion mechanism. In this scenario, a
first-order phase transition to strange matter or quark
matter triggers a second collapse of the proto-neutron star. If this collapse
is stopped before black hole formation and a second bounce occurs, a
second shock wave can be launched and expel the outer shells in a potentially
very powerful explosion \citep{fischer_18}. This scenario hinges on uncertain assumptions
about the nuclear equation of state, however, and its robustness and viability
is still under debate \citep{zha_21}.

\subsection{Parameter space for black-hole formation -- theory and observations}
Naively, one may expect black holes to form if and only if there
is no successful explosion. In reality, this is only a somewhat
useful approximation; black holes may sometimes form in successful
explosions as well as we shall discuss in Sections~\ref{sec:hne} and  \ref{sec:mass_loss}). Nonetheless, it is useful to first focus on the
question of successful shock revival first in order to approach
the systematics of black hole formation in core-collapse supernovae.

Detailed multi-dimensional radiation (magneto)-hydrodynamics simulations
have now matured to the point that many of them show successful shock
revival \citet{takiwaki_12,melson_15,lentz_15,mueller_17,mueller_19a,mueller_20,ott_18,burrows_19,burrows_20,vartanyan_22} and are able to produce explosion and remnant properties
broadly in line with observational constraints \citep{mueller_17,mueller_19a,bollig_21,bruenn_16}. This first-principle
approach is still of limited use for understanding the systematics of
the progenitor-remnant connection for two reasons. The immense computational
costs of self-consistent three-dimensional simulations only allow for a limited
exploration of the progenitor parameter space (mass, metallicity, multiplicity,
rotation). Only a few dozen such simulations have been performed so far by different groups.
Furthermore, first-principle simulations are still beset with
uncertainties and still cannot perfectly reproduce observational constraints
\citep{murphy_19}. For this reason, simpler phenomenological models with an appropriate
calibration and observations remain the most suitable means for determining
the progenitor-remnant connection and the parameter space for black hole
formation in particular.

Phenomenological models to determine the ``explodability'' of supernova
progenitors have so far exclusively considered the neutrino-driven scenario
for shock revival. A number of studies have used one-dimensional models
with various neutrino transport treatments and artificially enhanced
neutrino heating to study the parameter space for neutron star and
black hole formation, and in some cases the remnant mass distribution
as well \citep{oconnor_11,ugliano_12,ertl_15,ertl_20,sukhbold_16,ebinger_19,ebinger_20,ghosh_22}. The problem has also been studied using different (semi-)analytic approaches \citep{pejcha_15a,mueller_16a}. Large-scale parameter studies
are also possible in axisymmetry (2D) already \citep{nakamura_15}, but these
can only cover the initial phase of the explosion and the assumption
of axisymmetry severely impacts the dynamics of shock revival and of
the explosion for such models to be considered substantially superior
to the aforementioned approaches. Similarly, attempts to incorporate
multi-dimensional effects into one-dimensional simulations \citep{couch_20}
suffer from too many shortcomings to be considered a major improvement
\citep{mueller_19b} and are at odds with observational constraints (see below).

It must be borne in mind that phenomenological models need to
explicitly or implicitly incorporate calibration points or
constraints to predict the landscape of neutron star and black
hole formation. Common choices are to fix the explosion parameters to those
of SN~1987A \citep{ugliano_12} and possibly add extra constraints for
low explosion energies of the least massive supernova progenitors
\citep{ertl_15,sukhbold_16}. Other studies have used softer constraints such
as plausible limits on observed supernova explosion energies \citep{mueller_16a}. Further observational
constraints may inform the models, even if they are not explicitly incorporated. Considerable
care must therefore be taken to gauge the predictive value
of phenomenological supernova models. In some instances they
may furnish more of an interpretation or physically motivated
extrapolation from observations than firm theoretical predictions.

Despite the disparity of methods among phenomenological supernova models and the calibration uncertainties,
some robust features of the progenitor-remnant connection have nonetheless emerged.
The ``explodability'' of progenitors by the neutrino-driven mechanism
is strongly correlated with structural parameters of the stellar core
and its surrounding shells. A popular predictor for explodability
is the \emph{compactness parameter} $\xi$ \citep{oconnor_11}, which is defined as
\begin{equation}
    \xi_M=\frac{M/M_\odot}{R(M)/1000\, \mathrm{km}}
\end{equation}
where $M$ is a fiducial mass coordinate (measured in solar masses)
and $R(M)$ is the corresponding radius. Values of $\xi_M$ with $M$ in the range $1.75\texttt{-}2.5\,M_\odot$  have been found to provide
good proxies for explodability \citep{oconnor_11,ugliano_12}. The threshold value
of $\xi$ for black hole formation is subject to empirical calibration;
values of $0.2\texttt{-}0.45$ are commonly used \citep{oconnor_11,mueller_16a,aguilera_20}.

\begin{figure}
    \centering
    \includegraphics[width=\linewidth]{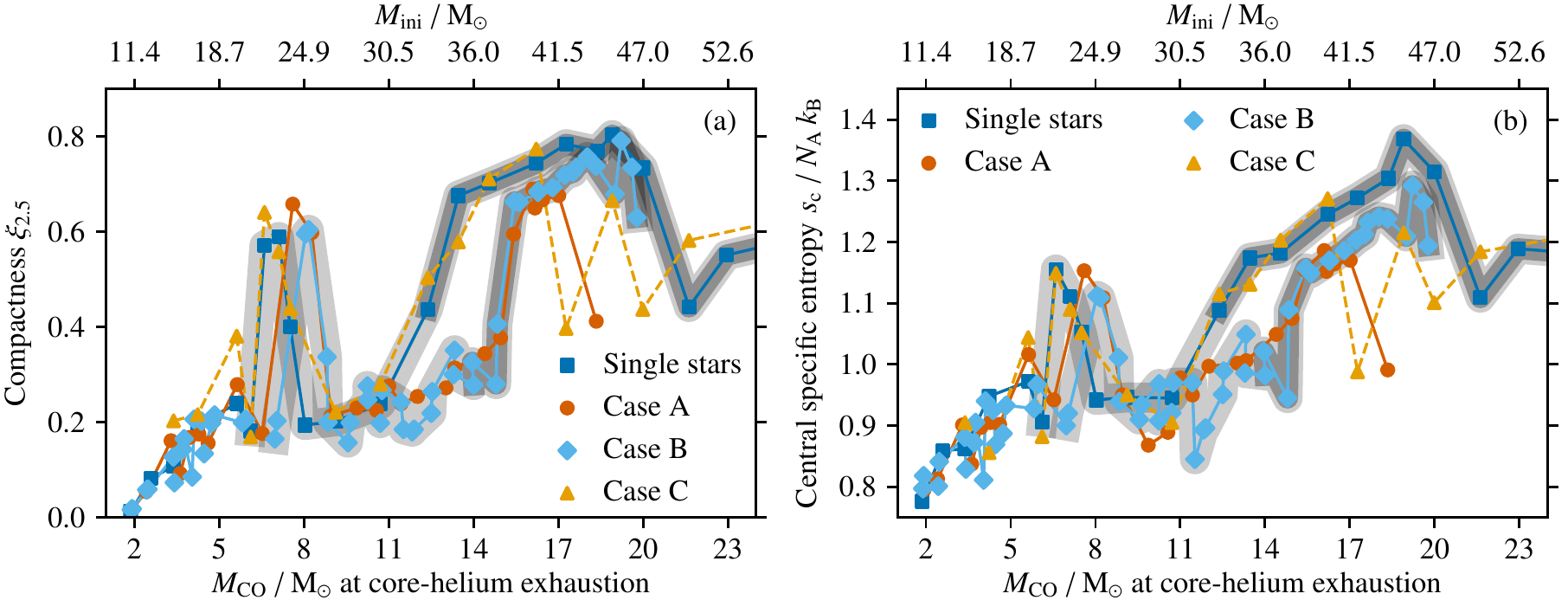}
    \caption{Core compactness $\xi_{2.5}$ (left) and central specific entropy, $s_c$ (right), as a function of CO core mass at helium exhaustion for single stars and stars that experienced mass loss in binaries  with different mass transfer cases (late Case~A near the end of main-sequence evolution, Case~B on or during the ascent to the red giant branch, Case~C after core He burning). The different C/O ratios after He core burning shift  the compactness at $7\texttt{-}8 M_\odot$ and also considerably reduces compactness  between
    $11 M_\odot$ and $15 M_\odot$ in He core mass in case
    of Case~A and Case~B mass transfer.
    Figure from
    Reference~\refcite{Schneider:2020},
    reproduced with permission from Astronomy \& Astrophysics, \textcopyright{ESO}.}
    \label{fig:co}
\end{figure}

While the compactness parameter has proved a popular measure for
explodability, it has no particular physical meaning to single it
out as a unique metric. Other structural parameters that are suitable
predictors for explodability also exist. The mass $M_4$ of the Fe-Si core
is also a good indicator for the outcome of core collapse, especially if combined with a second parameter
$\mu_4$ that essentially characterizes the density of the oxygen shell
\citep{ertl_15}, and can be linked to the mass accretion rate
onto the shock after the infall of the Si/O shell interface, which is often the point at which and explosion develops in detailed simulations.
A high binding energy of the shells outside the silicon-oxygen shell boundary
is also an indicator for black hole formation because
the binding energy tends to be strongly correlated with
both $M_4$ and $\mu_4$.  All of these core parameters tend to increase with initial stellar mass, the helium core mass after hydrogen burning, and the CO core mass after helium burning, although the dependence is not strictly monotonic (see Figure~\ref{fig:co}). Because of reduced mass loss, higher values of
$\xi_M$, $M_4$ and $\mu_4$ may be reached at collapse for a given initial mass at lower metallicity, especially for
initial masses $\gtrsim 20 M_\odot$
(cf.\ Figures~\ref{fig:MM1} and \ref{fig:MM3}).

\begin{figure}
    \centering
    \includegraphics[width=\linewidth]{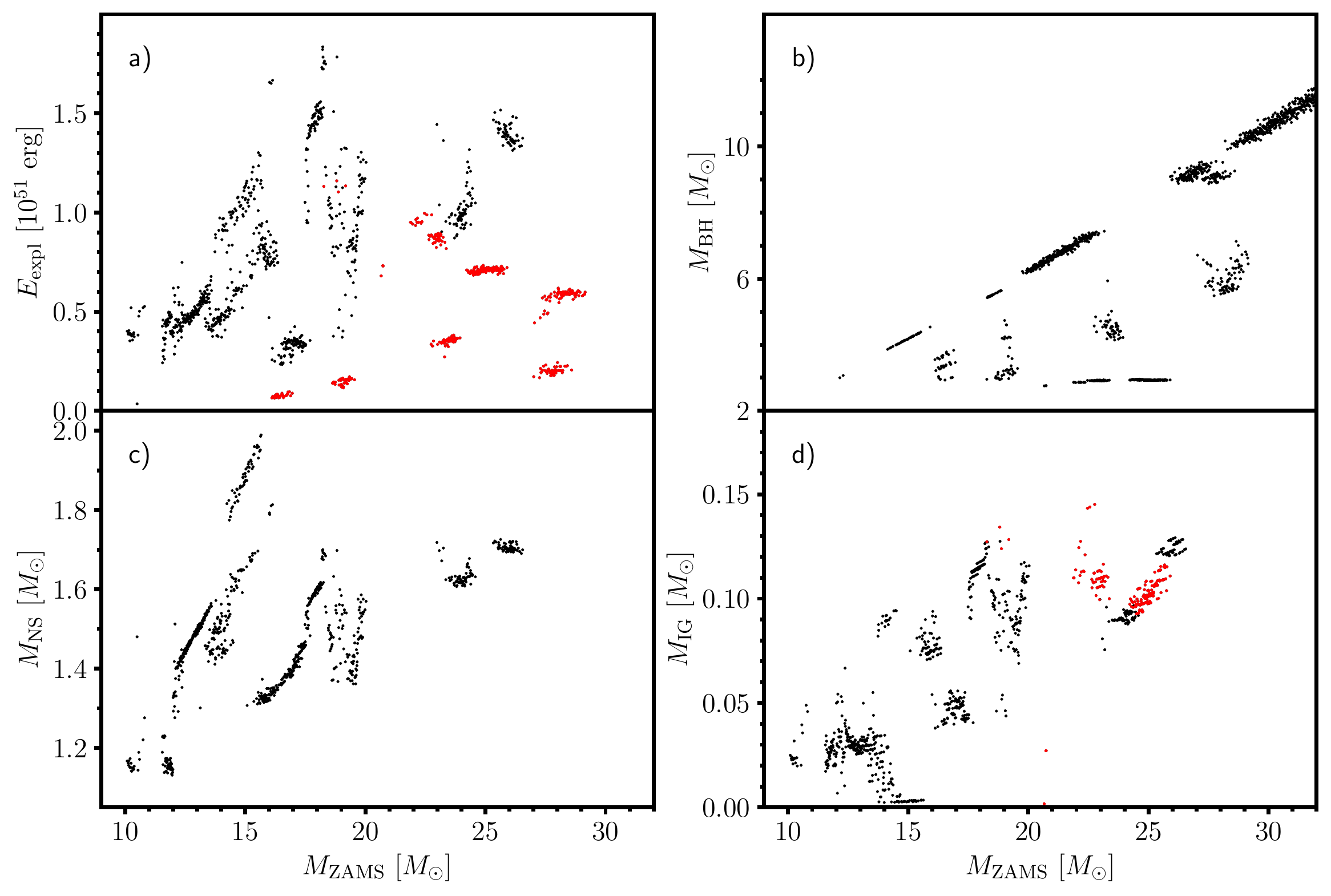}
    \caption{Supernova explosion and remnant properties for massive single stars of solar metallicity, based on an analytic explosion model including fallback\citep{mueller_16a,MandelMueller:2020}. The panels show a) the supernova explosion energy $E_\mathrm{expl}$ (if an explosion occurs), b) the black hole mass, c) the neutron star mass (depending on the type of remnant), and d) the ejected mass of iron-group (IG) elements, comprising mostly radioactive ${}^{56}\mathrm{Ni}$. Models that explode, but form black holes by fallback are indicated in red in panels a) and d).
    \label{fig:landscape}}
\end{figure}

Figure~\ref{fig:landscape} illustrates the predicted
outcomes of core collapse based on a study that is fairly
representative of the current phenomenological models.
For \emph{single-star} progenitors of solar metallicity, models usually predict
robust explosions as the final evolutionary stage of stars with birth masses
up to $18\texttt{-}20\,M_\odot$ \citep{oconnor_11,ugliano_12,sukhbold_16,mueller_16a}. At higher masses, they typically find black hole formation, often
interspersed with ``islands of explodability'' due to
non-monotonicities in the progenitor core structure as a function of stellar initial mass or, more generally, core mass (Section~\ref{par:C_shell}).  In particular, successful
explosions of single stars with birth masses
around $25 M_\odot$ have been found in
several studies \citep{ugliano_12,sukhbold_16,mueller_16a}. The underlying structural reason for the collapsing star to explode is a local
minimum in the compactness parameter and Fe-Si core size in
this mass range. This is due to the complex interaction of core masses left behind by each core burning phase and the sequence in which subsequent core and shell phases ignite, e.g., whether a critical mass for ignition or core contraction and collapse is reached.
This is somewhat similar to the concept of the Sch\"onberg-Chandrasekhar Limit\citep{SC42}, which defines a threshold mass for the isothermal core after hydrogen core burning on the main sequence, above which the core must contract and subsequently ignite helium core burning.  In particular, in Reference~\refcite{mueller_16a} this is related to mixing of the oxygen burning shell, located very close to the iron core, with the burning shells above (e.g., Ne and C burning), and the more volatile fuel being transported closer to the core and raising the entropy in these layers.
Fine details of the landscape of neutron star and black hole
formation are less robust. Some studies indicate the possibility
of islands of black hole formation as low as $15 M_\odot$ in
birth mass. At birth masses above $20 M_\odot$, the islands of explodability tend to appear in similar places regardless of the simulation methodology, even
though different models may disagree how big these islands are, or whether a particular island of explodability is present at all. This behaviour is easily understood by recognizing that the underlying explodability is set by the stellar structure, but
different phenomenological supernova models differ in the effective threshold (or ``water level'') for parameters like the compactness $\xi$ or Fe-Si core mass $M_4$.
When considering the final fate of massive stars as a function of \emph{initial} mass, one should bear in mind that the pattern of explodability also depends on mass loss by winds or eruptive mass loss events, and on binary interactions (see below), which may still open the possibility for successful explosions of stars with rather high birth masses well above
$20M_\odot$ in certain evolutionary scenarios.

Most phenomenological supernova models have focused on
solar-metallicity progenitors, but some have investigated
sets of stellar models of different metallicity \citep{oconnor_11,ebinger_20}, typically
including $Z=0$ and $Z=10^{-4}Z_\odot$. Due to reduced
progenitor mass loss, fewer (if any) islands of explodability
at high mass are predicted at these low metallicities compared
to the solar case.

For stripped-envelope supernova progenitors that have undergone
mass transfer in  \emph{binary stars}, the initial (pragmatic) assumption has
been that such progenitors will exhibit similar explodability as single-star progenitors
with the same helium core or C/O core mass \citep{Fryer:2012,VignaGomez:2018,MandelMueller:2020}. This leads to the expectation that
stripped-envelope progenitors with a C/O core mass of $\gtrsim 4.5 M_\odot$ will
mostly form black holes with the possibility of some interspersed instances of neutron star formation at higher values.
This was initially confirmed in a study\citep{ertl_20} that modelled stripped-envelope
progenitors as helium stars, although there were differences in detail between
helium stars and single stars, e.g., a higher threshold by
about  $1M_\odot$ in He core mass in the compactness peak
that corresponds to the first island of black hole formation.
A more nuanced picture (that also accounts for the shift in the compactness peak) emerges when considering the evolutionary
phase during which mass transfer occurs \citep{Schneider:2020}.
For Case~A mass transfer (during core H burning in the donor) and Case~B mass transfer
(after core H burning during the red giant phase or the ascent to the red giant branch), the dependence
of compactness on C/O core mass is indeed systematically shifted
(Figure~\ref{fig:co}) due to composition differences that are imprinted onto the core during core He burning. Since the He core can grow during He core burning due to hydrogen shell burning in single stars but not in stripped stars, He burning in single stars results in a higher O/C ratio\cite{brown_96} due to the extra supply of fresh $^4\mathrm{He}$, unless mass transfer occurs after He core burning has already finished.

In addition to a shift of the first compactness peak up by $\mathord{\sim} 1M_\odot$ compared
to the single-star case, there is also a much wider low-compactness valley at $8\texttt{-}14 M_\odot$ in
C/O core mass, giving rise to the possibility of successful
explosions up to high C/O core masses\cite{Antoniadis:2021}. On the other hand,
Case~C mass transfer (after the end of core He burning) results in a similar dependence
of explodability on the C/O core mass as in single stars. In terms of initial
masses, the windows for black hole formation for stars undergoing Case A or B mass transfer
may be considerably reduced; Schneider et al.\citet{Schneider:2020} report only a small birth mass range between
about $31 M_\odot$ and $35 M_\odot$ at solar metallicity as well as black hole formation at high
initial masses $\gtrsim 70 M_\odot$ for these channels.
As a result, the predicted merger rates for black-hole neutron star and black hole-black hole binary systems may be reduced by taking into account the structural effect of mass transfer on the pre-supernova structure\citet{Schneider:2020}.

There are, however, some noteworthy counterpoints to the aforementioned picture
in supernova theory. Some 3D simulations \citep{chan_18,chan_20,kuroda_18,ott_18,burrows_20,powell_20,powell_21} and tuned 1D simulations
\citep{couch_20} found \emph{shock revival} at high compactness. Shock revival does not guarantee
a successful explosion, however. If shock revival occurs in progenitors with massive cores,
massive oxygen shells, and very high compactness, the proto-neutron star will accrete
considerable mass after shock revival and likely form a black hole by fallback.
Consistent explosions in high-mass, high-compactness progenitors would also be at odds with observational
constraints (see below). Nonetheless, the possibility of shock revival in high-compactness
progenitors followed by black hole formation due to fallback needs to be considered
and will be highly relevant for the distribution of black hole masses, kicks, and
spins (Section~\ref{sec:mass_loss}).

The most direct \emph{observational} constraints on the parameter space for
neutron star and black hole formation come from the identification of supernova
progenitors. Progenitor masses have been inferred for a number of Type~IIP supernovae (hydrogen-rich supernovae with an extended luminosity plateau of about $100\, \mathrm{d}$)
from red supergiant progenitors by matching the brightness and colour of pre-explosion
images to stellar evolution tracks \citep{smartt_09}. The majority of these progenitors
will have evolved as single stars \citep{podsiadlowski_92,zapartas_21}. The inferred birth masses of the
progenitors are somewhat dependent on the treatment of convection during hydrogen
core burning in the underlying stellar evolution models\citep{sukhbold_14}, but there
is rather strong evidence that most red supergiants above $15\texttt{-}18 M_\odot$
do no explode \citep{smartt_09,smartt_15}. In view of possible alternative explanations for the lack
of high-mass red supergiant explosions and statistical uncertainties \citep{davies_20},
surveys for disappearing red supergiants have been suggested as a more direct
means to study the parameter space for black hole formation\citep{kochanek_08}. The observed
disappearance of a $25 M_\odot$ star \citep{adams_17,basinger_21} supports the hypothesis that
more massive progenitors mostly form black holes. Progenitor mass estimates
based on the nebular spectroscopy of Type~IIP supernovae is also consistent
with a lack of explosions at high progenitor masses \citep{jerkstrand_12}. Unfortunately,
the most readily available data -- supernova photometry -- cannot provide strong
constraints on the progenitor mass due to parameter degeneracies \citep{dessart_19}.

The evidence from well-studied historic supernovae and young supernova remnants requires
more careful interpretation because of various types of binary interaction.  In principle, matching the type of the compact remnants for historic supernovae to their progenitor or He core mass can help constrain the progenitor-remnant connection.
SN~1987A and the Cas~A supernova have left neutron star remnants
\citep{ho_09,page_20} and mass estimates place their progenitors at initial masses of $16\texttt{-}22 M_\odot$ for SN~1987A\citep{arnett_89} and $15\texttt{-}25 M_\odot$ for Cas~A\cite{young_06}. However, these numbers are based heavily on models, and in the case of SN~1987A, the attribution of a ZAMS
mass is questionable in the first place, as its
 progenitor  was likely the product of a late stellar merger.
 Mass loss definitely played a critical role in the progenitor evolution of Cas~A. By means of light echoes, Cas~A was
identified as a Type~IIb supernova\cite{krause_08} whose progenitor had undergone partial stripping
of the hydrogen envelope (possibly during the companion's supernova, which
would make the progenitor estimates questionable\citep{Hirai:2020}).
Their inferred helium core masses of $\sim 6M_\odot$
place them slightly below the first major island of black hole formation as predicted
by most phenomenological models. The young remnant W49B originated
in a stripped-envelope supernova (Type Ib/c) from a progenitor with an inferred
mass of $\mathord{\sim} 25M_\odot$,
and persuasive arguments have been made
that the explosion produced a black hole\citep{lopez_13}, which might point towards a collapsar engine, or towards fallback in an explosion driven by some other mechanism. Unfortunately, progenitor
mass estimates based on pre-explosion images for Type Ib/c supernovae are still scarce and cannot strongly constrain the parameter space for black hole formation among
massive stars that have undergone mass loss in binaries. For an identified progenitor system
for the Ib supernova iPTF13bvn, a tentative
mass estimate has been formulated based on binary evolution modelling and constraints from the photomoetry of the progenitor system and the supernova itself, putting the  pre-collapse
mass (i.e., final helium star mass) at about  $3.5 M_\odot$ \cite{bersten_14,eldridge_15}, but this estimate is not yet on par with those for Type~IIP supernovae.

Yet another, more indirect way to constrain dependence of
explodability on initial mass is to age-date the environments
of supernova remnants. Such age-dating of remnant environments
in M31 and M13 qualitatively supports the hypothesis
of missing explosions at higher mass \citep{jennings_11,jennings_14}.

\subsection{Collapsars, Hypernovae, and Gamma-Rays Bursts}
\label{sec:hne}
Black hole formation may be a crucial element in \emph{hypernovae}
with unusually high explosion energies up to $\sim 10^{52}\, \mathrm{erg}$ as opposed to the typical core-collapse supernova explosion energy of $\sim 10^{51}\, \mathrm{erg}$. Such events make up about $1\%$ of the supernova population in the local universe\citep{smith_11}
and possibly up to $10\%$ in low-metallicity environments
\citep{arcavi_10}. Starting with SN~1998bw\citep{galama_98}, it has been recognized that long gamma-ray bursts (GRBs) are associated with such hypernovae \citep{woosley_06}, although it is not clear whether all hypernovae  produce long GRBs.

The collapsar scenario\citep{macfadyen_99}, the characteristic features of hypernovae and gamma-ray bursts are explained by the formation of a black hole and accretion disk in the collapse of a rapidly rotating massive star. A non-relativistic wind outflow from the disk provides the energy of the hypernova and abundant radioactive $^{56}\mathrm{Ni}$ to power its light curve\citep{macfadyen_99}. The formation of the relativistic GRB jet likely involves the extraction of rotational energy from the black hole or the disk by magnetohydrodynamics effects
via the Blandford-Znajek\citep{blandford_77} or Blandford-Payne\citep{blandford_82} mechanism. To date, we still lack unambiguous observational evidence whether hypernovae and long GRB involve black hole formation, or whether rapidly rotating neutron stars
(``millisecond magentars'') are behind the relativistic jet
\citep{usov_92,duncan_92} and the hypernova explosion \citep{akiyama_03}.
It is also possible that similar disk-power engines akin to the collapsar scenario operate in some superluminous supernovae \citep{moriya_18b}.
An extensive review of current research on hypernovae and long GRBs as provided by recent reviews\citep{woosley_06,Janka:2012} is beyond the scope of this chapter. It is more pertinent to focus on how the collapsar scenario fits into the broader picture of stellar evolutionary channels to black hole formation and black hole birth properties.

After a black hole has formed in a rapidly rotating progenitor,
feedback from a collapsar-type engine will affect further accretion
onto the black hole roughly once the specific angular momentum $j$
of the infalling shells reaches the critical specific angular
momentum at the innermost stable circular orbit
($j_\mathrm{c,Kerr}\gtrsim 2/\sqrt{3}GM/c$ for a Kerr black hole,
$j_\mathrm{c,NR}\gtrsim 2\sqrt{3}GM/c$ for a non-rotating black
hole). Although the angular momentum of the black hole could
be small in principle when disk formation occurs, one usually
expects the black to have a high spin parameter $a=Jc/(GM^2)\approx 1$ based on actual stellar evolution models for hypernova progenitors
\citep{woosley_12}. Once an accretion-powered engine operates, the outflows will extract energy and angular momentum from the disk
and/or the black hole, and the feedback from the engine may quench
the accretion flow \citep{Batta:2019,Murguia:2020}. Qualitatively, one therefore expects a sub-population of black holes with lower mass and high spin parameter from rapidly rotating progenitor stars
\citep{Batta:2019}. The quantitative evolution of the black hole mass and spin parameter is somewhat more complicated and depends on how efficiently the outflows extract energy and angular momentum from the system. Depending on whether or not powerful magnetohydrodynamic jets form or not (which depends, e.g., on the field geometry in the accretion disk), the black hole may lose or gain energy (i.e., mass)
and angular momentum \citep{mckinney_12}.

Different scenarios have been proposed to account for progenitors with the requisite rapid rotation at collapse to enable hypernova explosions. In a scenario without binary interactions, sufficient angular momentum needs to be retained in stars that are born with rapid rotation.  Such rapid rotation could also be the result of binary star mergers during the pre-main sequence or on the main sequence.  Whereas without angular momentum transport, stars may easily reach critical rotation in their cores\citep{HLW00}, the big challenge in models is to retain sufficient angular momentum in their cores when angular momentum transport is considered.  The challenge includes the requirement that most pre-supernova stellar cores need to rotate slowly enough to be compatible with observed rotation rates of supernovae and supernova explosion energies\footnote{The rotational energy of a nascent neutron star with a rotation period of a few milliseconds would far exceed observed supernova explosion energies.}.  Next to transport of angular momentum, a key factor is loss of angular momentum due to stellar winds, in particular when the star has an extended red supergiant envelope with its huge specific moment of inertia.  One way around this is a stellar evolution path where the star already rotates very rapidly on the main sequence such that it remains fully mixed and undergoes what is called chemically homogeneous evolution \citep{HLW00, woosley_06a,yoon_06}, avoiding the red supergiant evolution phase.  The scenario still requires low metallicity to avoid loss of angular momentum due to extended Wolf-Rayet winds\cite{woosley_06a} and the contribution of magnetic stresses due to dynamo action \citep{spruit_02,heger_05} can be disfavourable to angular momentum retention.
This scenario, however, is not commonly realised in nature and the predicted GRB/hypernova fraction of less than $0.1\%$ at low metallicity\citep{yoon_06} appears too low to account for all observed events.  The conditions for chemically homogeneous evolution may also be reached due to spin-up by mass transfer in close-to-equal-mass binaries, with a likely break-up of the system after the supernova of the mass donor \citep{Cantiello:2007}.
Stellar mergers of evolved stars present a possible channel
to generate more rapidly spinning helium stars than can be easily produced in a single-star scenario \citep{podsiadlowski_04,fryer_05,woosley_06,Podsiadlowski:2010}.
This scenario is noteworthy because the ensuing collapsar would by construction give birth to an isolated black hole unless the progenitor system was a triple.
Proposed merger pathways include, e.g., the merger
of two helium cores of evolved stars during common-envelope evolution \citep{fryer_05}, or the merger of an evolved massive stars after He core burning with a low-mass star leading to
explosive ejection of the helium and hydrogen shell
\citep{Podsiadlowski:2010}.  Tidal spin-up of Wolf-Rayet stars in binaries, e.g., Reference~\refcite{Bavera:2021}, is discussed later in Sections~\ref{sec:HP_spin} and \ref{sec:binevol}.

\subsection{Pair-instability and pulsational pair instability supernovae.}
\label{sec:pisn}

\begin{figure}
    \centering
    \includegraphics[clip,viewport=48 100 688 536,width=\textwidth]{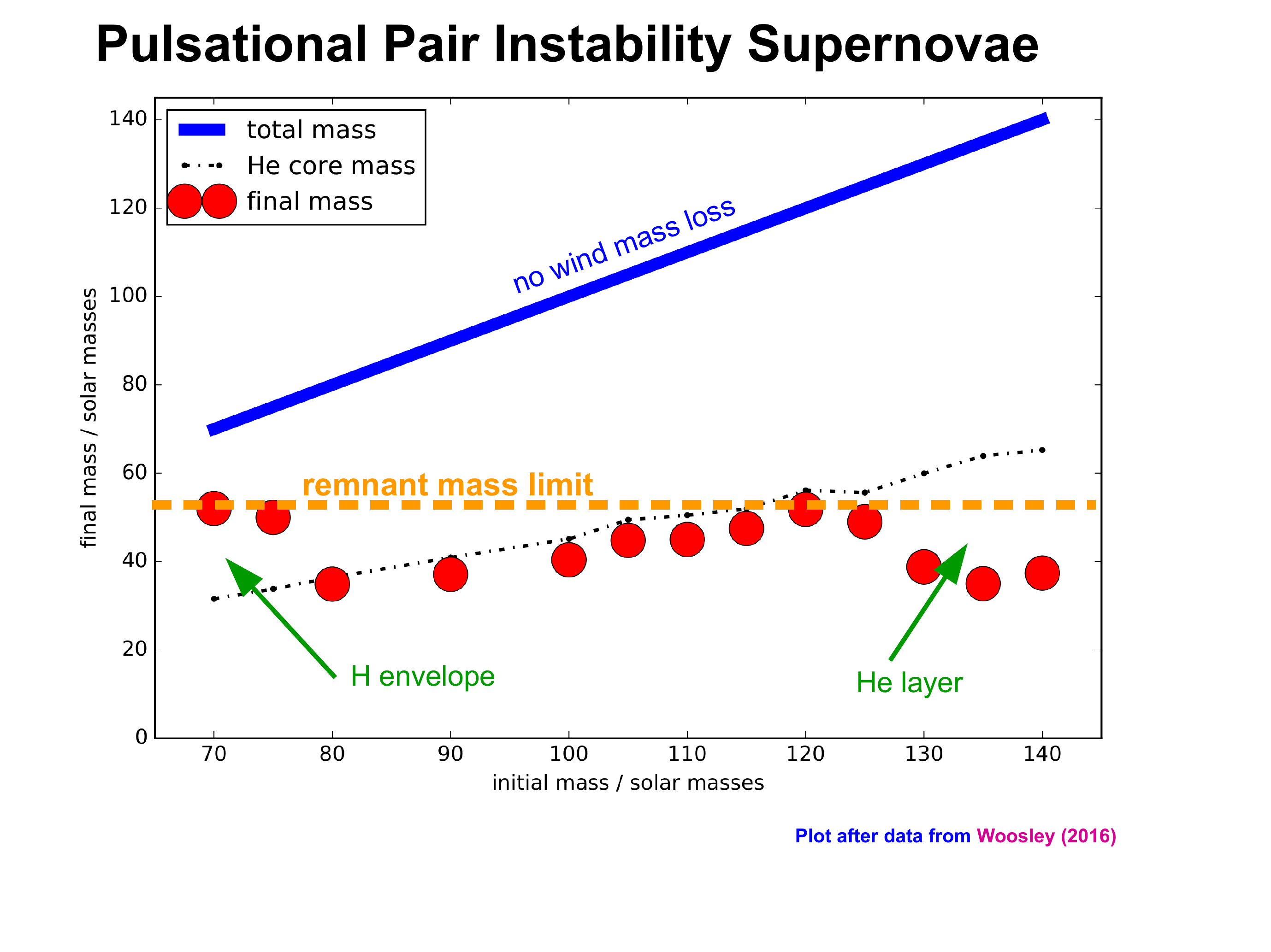}
    \caption{Final stellar masses of pulsational pair instability supernovae as a function of initial mass (\textsl{red dots}).  The \textsl{dash-dotted line} indicates the helium core mass.  Above $\mathord\sim80\,M_\odot$ in initial mass the entire hydrogen envelope is ejected, and above $\mathord\sim120\,M_\odot$ parts of the carbon-oxygen core are ejected as well.  The \textsl{solid blue line} shows the mass of the star in the absence of mass loss.  Pulsational pair instability continues to operate as long as the star is above a critical mass limit, or until a sufficiently large iron core is produced and the star collapses to a black hole.  The result is an effective upper mass limit for black holes that can be made this way (\textsl{orange line}).  Above $\mathord\sim140\,M_\odot$ the first pair-instability pulse is already powerful enough to entirely disrupt the star, and no compact remnant is left behind.   The data for the plot is taken from Reference~\refcite{woosley_17}; a somewhat different initial-final mass function for rotating very massive stars is found in Reference~\refcite{marchant_20}.
    \label{fig:ppsn}}
\end{figure}

When the mass of the helium core of a star at the end of core helium burning exceeds about $\gtrsim 40 M_\odot$, the star encounters the electron-positron pair creation instability \cite{fowler_64}.  Shortly after core carbon burning, which is radiative, the temperature exceeds $10^9\,$K and high-energy photons in the tail of the Planck spectrum can create electron-positron pairs.  The rest-mass of these pairs is taken from the internal energy of the gas, leading to a softening of the equation of state, i.e., lowering of the adiabatic index, $\gamma_\mathrm{ad}$, beyond the limit of stability\footnote{The star is stable when the condition is fulfilled.},
\begin{equation}
\int_0^M\frac{P(m)}{\rho(m)}\,\left(\gamma_\mathrm{ad}(m)-\frac43\right)\,\mathrm{d}m > 0\;,
\end{equation}
where $M$ is the total mass of the star, $P$ is the pressure, $\rho$ is the density,and  $m$ is the mass coordinate.    The star contracts rapidly on a dynamical timescale and encounters ``\emph{explosive}'' (very rapid) nuclear burning until the temperature has risen enough that new particles actually contribute sufficiently positively to the gas pressure.  The star re-expands and a shock wave forms that ejects the outer layers of the star.  After the pulse, the star cools down, contracts, and may encounter further pair instability pulses.  Subsequent stages of neon, oxygen, and silicon burning may be encountered in one or several pulses until a large enough iron core is formed and the star collapses to a black hole.  As the initial mass of the star and its post-helium-burning core increase, and also as one gets to higher pulse numbers in a sequence, the thermonuclear pulses get increasingly more energetic.  This leads to larger mass ejection, but also to larger post-pulse entropy in the core.  Larger entropies imply that the core takes longer to cool, with a longer Kelvin-Helmholtz timescale before the next pulse can occur.  There is, indeed, a critical transition when the post-bounce temperature drops much below $10^9\,$K: the gas becomes too cool to efficiently emit neutrinos, and has to cool by photon emission from the surface instead, resulting in inter-pulse cooling times as long as $10\mathord,000\,$yr\cite{woosley_17}.  In contrast, the inter-pulse phase between low-energy pulses may be as short as hours to days.  When the inter-pulse phase is of the order of years, for stars with initial mass of around $200\,M_\odot$, the timescale may just be right to produce bright outbursts from collisions of ejecta shells from subsequent pulses, as discussed below.

The production of neutron stars in the final collapse seems unlikely, and should be very rare at best.  The final collapse may yet trigger a final hypernova explosion or gamma-ray burst \cite{woosley_07} due to the collapsar mechanism\cite{woosley_93} (Section~\ref{sec:hne}).  The remarkable outcome of these pulses is that they imply an upper mass limit for black holes that can be made though this channel of around $\mathord\sim45\,M_\odot$\cite{woosley_17} (Figure~\ref{fig:ppsn}).

For initial stellar masses larger than $\mathord\sim140\,M_\odot$, i.e., helium core masses $\gtrsim65\,M_\odot$, even the first pulse is already energetic enough to entirely disrupt the star.  \emph{There is no compact remnant.}  This is the domain of the full pair instability  \cite{barkat_67,bond_84,heger_02} regime.  Kinetic explosion energies can range from $4\times10^{51}\,$erg to almost $10^{53}\,$erg, and the nucleosynthesis can range from being basically free from primary production of iron group elements to the production of more than $60\,M_\odot$ of $^{56}$Ni,\cite{heger_02} the same isotope that is responsible for powering the light curves of regular core collapse and Type Ia supernovae.  The extreme case is basically the equivalent of 100 Type Ia supernovae simultaneously in one spot.

At the high-mass end for full pair instability the entropy in the core gets so high that the $^{56}$Ni from silicon burning photo-disintegrates back into alpha particles, however, when the star explodes these recombine to $^{56}$Ni for the most part.
At the upper mass limit for full pair instability, i.e., for stellar masses of $\gtrsim260\,M_\odot$, helium core masses $\gtrsim130\,M_\odot$, the entropy in the core gets so high that even the alpha particles photo-disintegrate into free nucleons.  These photo-disintegrations take out internal energy of the gas, reducing the pressure and thereby softening the equation of state similar to the pair instability.  In particular, the instability caused by photo-disintegration of the alpha particles is so strong that the collapse is not turned around by an outward shock, and instead continues as a direct collapse to a black hole.  Unless there is sufficient angular momentum to form an accretion disk, the entire star should collapse.  The resulting black holes should have initial masses of at least $130\,M_\odot$ for pure helium cores\cite{heger_02}, larger for larger stars or if the hydrogen envelope was not lost.  If there is sufficient angular momentum in the star at collapse, the collapsing core may become a strong source of gravitational waves\cite{fryer_01}, or the angular momentum centrifugal barrier may generate an accretion disk powering a long-duration gamma-ray burst, leading to more mass ejection and reducing the black home mass (see Section \ref{sec:hne}).

The consequence of the upper mass limit for black holes from pulsational pair instability ($45\,M_\odot$) combined with the lower mass limit for black holes beyond the full pair instability regime ($130\,M_\odot$) is a gap in black hole birth mass function\cite{belczynski_16}.  This straight-forward prediction has been challenged by gravitational-wave observations that imply likely detections of black holes with masses within this mass gap\cite{GWTC3:pop}.
Suggested solutions for the existence of mass-gap black holes include low-metallicity stars just below the pair instability supernova limit that may be as massive as $70\,M_\odot$ at the time of core collapse\cite{belczynski_20} or modifications to stellar physics such as key nuclear reaction rates\cite{farmer_20,woosley_21}, binary evolution, rotation, and accretion after black hole formation\cite{vanSon:2020,SafarzadehHaiman:2020,Tagawa:2021,woosley_21}.  Whereas these may be able to shift, and, in part, even narrow the mass gap, these works show that eliminating the gap entirely remains a challenge.  Another possibility is that the merging black holes are products of earlier mergers, perhaps through dynamical formation in globular or nuclear clusters\cite{Rodriguez:2019,Yang:2019}.

Transient observations have so far been unable to provide further insights on the  predicted pair instability mass gap.
It is noteworthy, though, that there is no unambiguous detection of a pair instability supernova yet. Due to the large ejected mass of $^{56}\mathrm{Ni}$, pair-instability supernovae were adduced as an explanation for superluminous supernovae early on
\citep{gal-yam_09}, but the observed events tend to differ markedly from model predictions in terms of their light curves (especially in terms of the rise time before peak) and spectra \citep{kasen_11,dessart_13,jerkstrand_16}.
Different from pair-instability supernovae proper, there is a considerable class of observed transients that fit the characteristics of pulsational pair-instability supernovae.
Whereas the individual pulses may not be particularly powerful in terms of kinetic energy compared to ordinary supernovae, collisions of shells from different pulses can produce very bright transients \citep{woosley_07} due to the high efficiency of conversion of kinetic energy to observable photons.  This suggests that some
\emph{superluminous supernovae} might be pulsational pair instability events. Depending on the mass, metallicity, rotation rate, and prior mass loss history, a wide variety of light curves can be produced, and both hydrogen-rich Type~IIn superluminous supernovae (with evidence for interaction in the form of narrow emission) and Type~I  superluminous supernovae can be accounted for \citep{chatzopoulos_12,moriya_13,yoshida_16,woosley_17}.
Observationally, the narrow emission lines in Type~IIn superluminous supernovae constitute strong evidence that these are interaction-powered \citep{moriya_18}. Bumps and undulations in the light curves of both Type~II and Type~I superluminous supernovae\citep{nicholl_15a,nicholl_16,inserra_17,nyholm_17}
 can also be interpreted as signs of interaction,
although alternative interpretations are often possible.
In many cases, high ejecta masses have been inferred for such interacting superluminous supernovae from light curve fitting
\citep{woosley_07,moriya_13,nicholl_15,tolstov_17} and
nebular spectroscopy\citep{jerkstrand_17}, which is compatible with the pulsational pair-instability scenario. The progenitor masses cannot be determined sufficiently well to verify the nature of the progenitors, however, let alone to constrain the mass range for the pulsational-pair instability channel.

\section{Black holes at birth: masses, kicks and spins}
\label{sec:birth_properties}

\subsection{Amount of mass loss during collapse}
\label{sec:mass_loss}
The dependence of ``explodability'' on stellar mass, rotation, and metallicity
is only one ingredient for understanding the observed population of stellar-mass
black holes. It has been recognized that in many instances of black hole
formation, partial mass ejection is likely to occur, which has important
implications for the birth distribution of black hole masses. The possibility
of partial mass ejection is most evident in the case of the collapsar scenario
(Section~\ref{sec:hne}). However, the impact of partial mass ejection on black hole
birth parameters has been studied more extensively for other scenarios.

Already in the 1980s it was suggested\citep{Nadyozhin:80} that massive stars may
eject part of their envelope after iron core collapse even if the shock is never
revived. The energy loss through neutrinos during the proto-neutron star phase
reduces the gravitational mass of the star, which disturbs the hydrostatic
equilibrium in the envelope. As a result, a sound pulse is launched, which
may eject tenuously bound envelope material. This idea, known as
Nadyozhin-Lovegrove mechanism, has been developed further in
recent years using numerical simulations \citep{Lovegrove:2013,fernandez_18,ro_19} and
analytic theory for the wave pulse launched by the reduction of the gravitational
mass \citep{coughlin_18,linial_21}. For plausible assumptions about the energy loss through neutrinos
and the black hole formation timescale, the hydrogen envelope is likely to
be ejected in the case of red supergiant progenitors \citep{fernandez_18}. For blue
supergiants with more compact envelopes, only $\sim 0.1 M_\odot$
will be lost, very little mass loss is expected for Wolf-Rayet stars,
and scenarios without mass ejection are also conceivable \citep{fernandez_18}. Mass
ejection due to this mechanism would give rise to a long-lived
red transient with a small energy of $\mathord{\lesssim}10^{48}\,\mathrm{erg}$ \citep{Lovegrove:2013,lovegrove_17,fernandez_18},
and may be followed by faint emission due to fallback for up to several
years \citep{coughlin_18}.  In addition,
there will be a brighter and bluer luminosity peak from shock breakout
ov $3\texttt{-}70\, \mathrm{h}$ \cite{lovegrove_17}.  Observations have yet to positively identify such a transient from the shedding of the envelope.

To approximately account for mass ejection by the Nadyozhin-Lovegrove
mechanism, recent phenomenological supernova models often assume that
the black hole mass will be given by the hydrogen-free mass of the progenitor,
although this may underestimate the black hole mass in some cases.

Genuine fallback supernovae present a more complicated case of partial
mass ejection. In fallback supernovae, the shock is successfully revived,
but the (proto-)neutron star eventually accretes enough mass later on to collapse
to a black hole. Several fallback scenarios can be distinguished.
In the case of early fallback, continuing accretion after early shock
revival already drives the neutron star to collapse during the first seconds to
minutes of the explosion. In the case of late fallback, accretion onto the neutron star is initially quenched,
but some of the ejecta fall back after they are decelerated
by one of the reverse shocks that form when the forward
shock runs across a shell interface \citep{fryxell_91,hachisu_90,herant_91}. In explosions of
red supergiant progenitors, a strong reverse shock forms when the forward
shock crosses the helium/hydrogen interface, transiently accelerates and then decelerates
again as it scoops up more material from the hydrogen envelope.
Nonetheless, fallback by deceleration
in the reverse shock usually adds little mass onto the remnant.
Fallback masses were mostly limited to $\lesssim 10^{-2} M_\odot$ in phenomenological
1D supernova models of solar-metallicity single-star progenitors
\citep{ertl_16b}.

Fallback can become much more dramatic, however, when the energy input
by the supernova engine exceeds the binding energy of the outer shells
only by a moderate margin \citep{chan_18,chan_20}. In this case, considerable fallback can occur
already during the early phase of the explosion because accretion downflows
are not quenched after shock revival \citep{chan_18,kuroda_18,ott_18}. At later stages, more
fallback can occur as the forward shock and the matter in its wake are slowed
down without the need to involve a reverse shock \citep{chan_18,chan_20}. The mechanisms
governing the final explosion energy and fallback mass in such marginal explosions are only qualitatively understood at this point. After the supernova
engine has stopped, the initial energy of the blast wave will be drained
as the shock scoops up bound material. Once the shock becomes sufficiently
weak, it will turn into a sonic pulse that transports energy through the
star without transporting matter, and little further energy will be lost
from the pulse \citep{chan_18,chan_20,stockinger_20}. The final mass cut is set roughly by the
point where the shock leaves the weak shock regime
again (i.e., post-shock Mach numbers reach $\gtrsim 1$)
 as it proceeds
to shells with smaller sound speed \citep{chan_18}. For a marginal explosion
to succeed, black hole formation must not occur too early, however. If
the black hole is formed when the shock has not crossed the sonic point
of the infall region yet, the incipient explosion is likely completely
stifled\citep{powell_21}, and only part of the envelope may be ejected by
the Nadyozhin-Lovegrove mechanism.

Multi-dimensional effects are extremely important in marginal explosions
with early fallback. So far only a handful of multi-dimensional simulations
of fallback after black hole formation have been conducted\citep{chan_18,chan_20,stockinger_20}; these have
been helpful for identifying the aforementioned principles. They also
showed that fallback in marginal explosions can produce black holes
over a considerable mass range from close to the maximum neutron star mass
to almost complete collapse \citep{chan_20}. In particular, fallback can explain
entities like the $2.6 M_\odot$ compact object in the merger event
GW190814\citep{GW190814}. Estimating the effect of fallback on black hole
populations using phenomenological 1D supernova simulations or analytic models
is more difficult. Current models show considerable variations in the
fraction of stars affected by strong fallback \citep{ertl_16b,mueller_16a,MandelMueller:2020,antoniadis_22}. There is, however,
agreement that fallback will produce a sizable number of low-mass black
holes and populate the ``mass gap'' that was formerly assumed to exist
between at $2\texttt{-}5 M_\odot$ between the most massive neutron stars
and the least massive black holes in known X-ray binaries\citep{Ozel:2010,Farr:2010}.
The amount of ejected mass in the case of a partially successful
explosion of a rapidly rotating progenitor
(which includes the collapsar scenario) is less well understood.

Aside from GW190814, there is additional circumstantial evidence
for partial mass ejection after black hole formation.  Dark objects with masses between
$2\texttt{-}5 M_\odot$ have been observed in microlensing
experiments\citep{WyrzykowskiMandel:2019,Mroz:2021} and in some detached binaries\citep{Thompson:2019}.
Abundances in some ultra metal-poor stars that were likely polluted by
one or a few supernovae can best be understood as resulting
from the removal of the iron group elements and sometimes
some intermediate mass elements from the inner ejecta of a supernova \citep{nomoto_06,keller_14}.
The composition of the black-hole companion in
Nova Scorpii also suggests pollution by a fallback supernova
that formed the black hole \citep{podsiadlowski_02}. Evidence from transient observations
is more dubious. Suggestions that faint Type IIP supernovae
with small ejected mass of ${}^{56}\mathrm{Ni}$ are black-hole
forming fallback events proved unlikely upon more recent analysis
\citep{spiro_14}. Some superluminous supernovae may be interpreted
as being powered by fallback \citep{moriya_18a,moriya_18b}, but a smoking gun
for this interpretation is lacking.

\subsection{Asymmetry / kicks?}
The multi-dimensional nature of the modern fallback scenario
opens up the possibility of strongly asymmetric mass ejection,
which could result in sizable black hole kicks due to
momentum conservation. Initial analytic estimates
envisaged the possibility of similar kick velocities
for black holes and neutron stars \citep{janka_17}. Only two
3D simulations have yet addressed black-hole kicks from
fallback and present a more nuanced picture with a kick of
$\mathord{\sim} 500\, \mathrm{km}\,\mathrm{s}^{-1}$ (i.e.,
slower than the fastest neutron star kicks) for a case
with moderate fallback, and a relatively
small kick of tens of $\mathrm{km}\,\mathrm{s}^{-1}$ for strong fallback in a weak explosion.
The physics evinced by these
simulations suggests that high-velocity black hole kicks
are confined to black holes of relatively low mass.
Attempts to extrapolate these results to black hole
populations involve some calibration of the absolute
scale of black hole kicks. However, theory does point
to a less populated
high-kick tail and a more pronounced low-velocity
peak in the black-hole kick distribution compared to
the neutron star kick distribution with its peak
at non-zero kick velocity.
Even in case of complete fallback, asymmetric neutrino emission prior to black hole formation may still impart kicks of
a few $10\, \mathrm{km}\, \mathrm{s}^{-1}$ onto
the black hole \citep{rahman_22}, although these
estimates are still based on 2D simulations that may somewhat overestimate the emission anisotropy.

The observational evidence for black hole kicks is somewhat uncertain.  It appears that heavier black holes such as Cygnus X-1 were born with very low kicks of $\lesssim 10$ km s$^{-1}$ \citep{Neijssel:2020CygX1}.  On the other hand, lighter black holes have been inferred to have larger kicks of $\gtrsim 200$ km s$^{-1}$ through observations of the positions and velocities of BH low-mass X-ray binaries (see section \ref{sec:binevol}) \citep{Repetto:2017,Atri:2019} (though see \citep{Mandel:2015kicks}), with further evidence potentially provided by observed spin-orbit misalignments in BH binaries \citep{Poutanen:2021}.

\subsection{Black hole spin}
\label{sec:HP_spin}

In the simplest picture of partial mass ejection, the black hole
birth spins will simply be determined by the amount of total
angular momentum in the progenitor star inside the mass cut
(envisaged as a spherical demarcation line between the black
hole and the ejecta). In the case of rapidly rotating progenitors,
this is likely a good approximation.

It is not, however, trivial to form rapidly rotating black holes.  Whereas typical initial stellar rotation is easily sufficient to make Kerr black holes, the amount of angular momentum actually present in the core largely depends on how much of the initial angular momentum is retained to core collapse.  This, in turn, strongly depends on \textsl{i)} angular momentum loss from the surface of the star and \textsl{ii)} the transport of angular momentum.  Magnetic fields may be quite efficient in angular momentum transport\cite{heger_05}; angular momentum loss through winds is also efficient except for low metallicities\cite{woosley_06a}.  Co-rotation or spin-up by accretion in close binaries in early evolutionary phases is still subject to angular momentum losses later in the evolution.  Some black hole progenitors, however, may be placed in very close binaries after the envelope is removed, e.g., by a common-envelope event, so that the naked helium core is efficiently tidally spun up and remains rapidly spinning until core collapse, as discussed in Section~\ref{sec:binevol}.  While this may limit angular momentum loss from the surface, the spin-up of the star has to occur early enough during the evolution to still impart enough angular momentum onto the very core - as soon as the critical angular momentum for Kerr black holes corresponds to surface rotation rates in excess of Keplerian rotation for a rigidly rotating star, sufficient spin-up of the core is no longer possible.

For slowly rotating progenitors, the black hole can, however,
also be spun up by asymmetric fallback. Due to the large lever
arm, relatively small amounts of fallback and small non-radial
velocities can impart significant angular momentum onto the
black hole. A recent 3D simulation showed that black hole
spin parameters of  $\mathord{\sim}0.25$ are within reach
for low-mass black holes ($\mathord{\sim}3 M_\odot$)
\citep{chan_20}. Spin-up of
black holes by asymmetric fallback will undoubtedly exhibit
large stochastic variations, and more systematic theoretical
and computational studies are required to predict its effect
on the distribution of black hole spins.

The spin of a black hole changes the radius of the innermost stable circular orbit, which sets the inner radius of the accretion disk.  Therefore, spins of accreting stellar mass black holes can be inferred through the observations of continuum X-ray flux or reflection lines from the accretion disk.  Spin measurements suggest a broad distribution of spins, from nearly non-spinning to nearly maximally-spinning \citep{MillerMiller:2015,Reynolds:2020}, although all inferred spins are model-dependent, so caution is warranted (see  Chapter IV of this volume).  Gravitational-wave observations (see Section \ref{sec:binevol} and Chapter VIII of this volume) also allow spins to be inferred, albeit with limited precision.  There is some debate in the literature regarding the spin distribution of merging BH binaries \citep{GWTC3:pop,Roulet:2020,Galaudage:2021,Callister_22}; it is possible that some merging binary BHs have negligible spins, while others have moderate combined spins preferentially aligned with the direction of the orbital angular momentum.

\section{Black holes in binaries}
\label{sec:binevol}

\begin{figure}
    \centering
    \includegraphics[width=\linewidth]{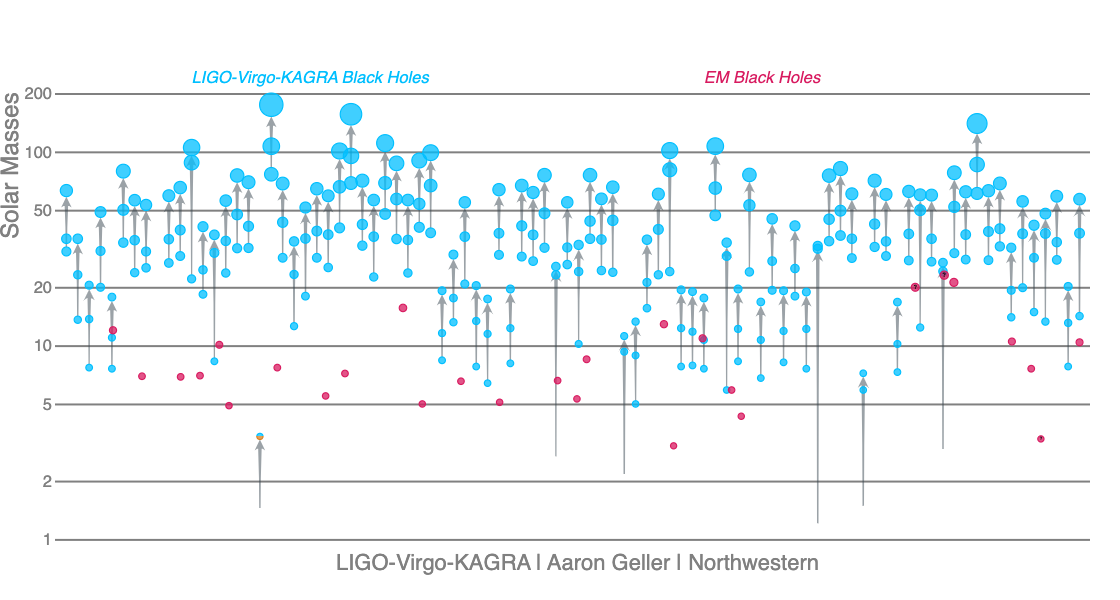}
    \caption{The masses of black holes measured through gravitational waves (blue, in chronological order, with merger product masses shown) and electromagnetic observations of X-ray binaries and detached binaries (red, random order).  Figure courtesy of Aaron Geller, Northwestern University and LIGO--Virgo--KAGRA collaborations.}
    \label{fig:graveyard}
\end{figure}

Most stellar-mass black holes are observed in binaries, through either mass transfer from a non-degenerate companion onto the black hole (X-ray binaries), detached binaries in which the presence of the black hole is inferred through the orbital motion of the optical companion, or mergers observed via their gravitational-wave signature.  Moreover, massive stars that go on to form black holes are typically born in binaries or systems with even more companions \citep{MoeDiStefano:2017}.  Black hole progenitors in such systems frequently gain mass from their companion, experience mass stripping by the companion, or are tidally spun up by the companion \citep{Sana:2012}. These interactions play a crucial role in black-hole formation and in determining the mass and spin of the black hole.  Consequently, even the single black holes, particularly those observed via gravitational micro-lensing \citep{WyrzykowskiMandel:2019,Mroz:2021} likely came from binaries and experienced binary interactions \citep{AndrewsKalogera:2022}.  With the exception of microlensing observations, binaries are also responsible for all mass measurements of stellar-mass black holes, as shown in Figure \ref{fig:graveyard} and further discussed in Chapter IV of this volume.  In this section, we briefly summarise the theoretical models of the impacts of binarity on the formation and properties of black holes.

We can broadly divide binary interactions into two categories: those that happen prior to BH formation and those that take place after the BH is formed.

Prior to BH formation, the key interaction mechanisms are mass transfer and tides.

If either star in a binary is going to make a black hole, it is likely to be the primary, the initially more massive star, though the secondary may also follow.  Mass ratio reversal through mass transfer can lead to the formation of, say, a neutron star from the primary and a black hole from the initially less massive secondary, but such outcomes are expected to be rare \citep{Debatri:2020}.  As the primary expands at the later stages of main sequence evolution or, particularly, after evolving off the main sequence, tidal gravity from the secondary will increasingly distort it and may ultimately lead to mass transfer (Roche lobe overflow).  This will reduce the star's mass and possibly alter its structure and even affect future collapse outcomes \citep{Schneider:2020}.  Mass transfer that removes the primary's envelope, which contains the bulk of its angular momentum, may leave behind a relatively slowly rotating core \citep{FullerMa:2019,Belczynski:2020,MandelFragos:2020}.  A stripped star may also experience particularly strong Wolf-Rayet winds, driving further mass loss and spin-down.  Of course, mass transfer can also add mass to the companion and change the orbital separation, potentially impacting future interactions.

In some cases, mass transfer may become dynamically unstable, leading to a common-envelope phase \citep{Ivanova:2013}, which may in turn result in a stellar merger.  The merger product may still form a black hole, but perhaps one with unusual features, such as a non-standard ratio of the core and envelope masses; such systems may, for example, explain (pulsational) pair instability supernova candidates and unexpectedly large black hole masses in high-metallicity environments \citep{VignaGomez:2019, Costa:2022}.

Tides primarily impact orbital evolution, synchronising and perhaps circularising the binary.  In the process, they can spin up the BH progenitor.  While that angular momentum may be removed with the envelope, in some cases, tidal spin-up may accompany mass transfer in such a way that the envelope is removed and the remaining core is spun up at the same time, possibly explaining some rapidly spinning BHs such as Cygnus X-1 \citep{Qin:2018, Neijssel:2020CygX1}.

Rapid rotation through tidal coupling may also cause efficient circulation within the main-sequence BH progenitor \citep{Eddington:1925,Sweet:1950}.  This could lead to  mixing throughout the star and, ultimately, chemically homogeneous evolution, in which the entire stellar mass of hydrogen fuses into helium, not just the core \citep{EndalSofia:1978,Heger:2000}.  This process may lead to the formation of close pairs of black holes from over-contact binaries \citep{MandelDeMink:2016,Marchant:2016}.

Binarity may also be important during the formation of the BH itself if some of the stellar material is torqued by the companion before falling back into the BH, thus contributing to BH spin, though this process may rely on some fine tuning \citep{Batta:2017,Schroeder:2018}.

Following BH formation, mass transfer, winds, and possibly gravitational-wave emission and/or dynamical interactions become key.  Tides do not affect the BH after its formation, though they can still impact subsequent binary orbital evolution, and could be responsible for the tidal spin-up of the secondary (which could go on to become another black hole \citep{Kushnir:2016,Zaldarriaga:2017,Belczynski:2020,Bavera:2020}).

Mass transfer onto a black hole may allow for electromagnetic observations of BHs as the accreting material radiates in X-rays.  BH X-ray binaries (XRBs) are generally divided into low-mass and high-mass XRBs, with the former fed by Roche lobe overflow from a low-mass companion and the latter by winds from a high-mass companion.

Low-mass XRBs can be very long-lived (on timescales of Gyrs), though often transient in nature as mass accretion stops and re-starts.  The total mass reservoir, however, is sufficiently small that the BH is unlikely to accrete a lot of mass or spin, although some studies suggest that such low-mass XRBs may be the evolutionary outcomes of intermediate-mass XRBs with significant mass accretion and BH spin-up \citep{Podsiadlowski:2003,FragosMcClintock:2015}.

Meanwhile, high-mass XRBs are necessarily short-lived, with lifetimes of order a Myr or less due to the short lifetime of massive stars.  This is much less than the mass doubling time of a black hole accreting at the Eddington limit ($\gtrsim 30$ Myr), and hence the BH cannot accrete a significant fractional amount of mass or get appreciably spun up \citep{KingKolb:1999} (but see Refs.\cite{vanSon:2020,ZevinBavera:2022} for a discussion of the impact of super-Eddington accretion).

In recent years, several black holes have also been observed in non mass-transferring binaries \citep{Thompson:2019, Giesers:2019}, with prospects for future detections through Gaia data \citep{Chawla:2021}.  However, observations of such detached BH binaries are notoriously challenging, with several recent candidates including LB-1 \citep{Liu:2019}, the `unicorn' \citep{Jayasinghe:2021} and the `giraffe' \citep{Jayasinghe:2022} ruled out by subsequent re-analyses \citep{Eldridge:2020,Shenar:2020,AbdulMasih:2021,ElBadry:2022}.

Our discussion so far has focussed exclusively on isolated binaries.  However, additional channels for BH binary formation and evolution include dynamical interactions in addition to stellar and binary evolution. Hierarchical triple systems may experience Lidov-Kozai oscillations leading to enhanced inner binary eccentricity \citep{Lidov:1962,Kozai:1962}. Interactions in dense stellar environments such as globular and nuclear clusters can introduce BHs into binaries through replacements and subsequently tighten those binaries \citep{Sigurdsson:1993,Kulkarni:1993}.  Similar dynamical interactions of stellar-mass black holes in AGN discs additionally include the possibility of significant accretion onto these BHs \citep{Bellovary:2016,Tagawa:2020}.

If two black holes are sufficiently close, gravitational-wave emission can drive them to merger.  The timescale for this merger scales with the fourth power of the separation and the inverse cube of the mass \citep{Peters:1964}, so that the 30 $M_\odot$ BHs that merged in the first gravitational-wave detection GW150914 \citep{GW150914} would have needed to be separated by less than a quarter of an astronomical unit in order to merger within the current age of the Universe.  Almost a hundred merging binary BHs have been detected as of 2022, providing the largest catalog of known stellar-mass BHs \citep{GWTC3,GWTC3:pop}.  The implications of these observations for the formation and evolution of stellar-mass black holes are the topic of very active ongoing work, partly summarised in a set of recent reviews \citep{MandelFarmer:2018,Mapelli:2021,MandelBroekgaarden:2021}; see  Chapter VIII of this volume for more details.

\section{Concluding Remarks}
\label{sec:summary}

A complete end-to-end understanding of the physics of stars and eventually the properties of black holes they make requires full three-dimensional simulations covering the full range from the micro-physics to the integral spatial and temporal scale of the stars,   Such calculations do not seem feasible for the foreseeable future based on current technology.  Practical approaches at our disposal include work toward understanding the physical processes to the extent that they can be accurately modelled on long time-scales, in lower dimensions, allowing to replace resolved microphysics by sub-grid models to be included in simulations on the integral scale of the problem -- the stellar scale -- or even on the system scale for multi-star studies.

The advances in computational modelling will thrive on insights obtained from the growing stream of observations in multi-messenger astronomy that combines, e.g., gravitational waves, cosmic rays and high energy neutrinos, and the entire range of electromagnetic observations over a vast range of time-scales, from sub-second transients to years-long light curves, to quasi-steady objects preceding violent stellar deaths.  New techniques for data analysis, such as machine learning or, in the long run, quantum computing, may allow us to better exploit the data to constrain theoretical models.
Genuinely new approaches are needed to better address key problems such as, e.g., black hole formation in supernovae and mass transfer in binaries.  We require a deeper understanding of how the physics of single and binary star evolution and supernova explosions connect to each other.

In the near future, a challenge lies in understanding the formation of black holes from the first generation of stars, where direct observations of the individual objects remain a challenge except for a few rare circumstances of chance magnification due to strong gravitational lensing\cite{welch_22} and caustic crossings with their huge magnifications on the order of $10\mathord,000$ \cite{windhorst_18}.

\section*{Acknowledgements}

A.H.\ was supported by the Joint Institute for Nuclear Astrophysics through Grant No.~PHY-1430152 (JINA Center for the Evolution of the Elements) and by the Australian Research Council (ARC) Centre of Excellence (CoE) for Gravitational Wave Discovery (OzGrav) through project number CE170100004, and by the ARC CoE for All Sky Astrophysics in 3 Dimensions (ASTRO 3D) through project number CE170100013.
B.~M.~acknowledges support by ARC Future Fellowship FT160100035.
I.M.~acknowledges support from the Australian Research Council Centre of Excellence for Gravitational  Wave  Discovery  (OzGrav), through project number CE17010004.  I.M.~is a recipient of the Australian Research Council Future Fellowship FT190100574.  Part of this work was performed at the Aspen Center for Physics, which is supported by National Science Foundation grant PHY-1607611.  I.M.'s participation at the Aspen Center for Physics was partially supported by the Simons Foundation.

\bibliographystyle{ws-rv-van}
\bibliography{bibliography,Mandel}

\end{document}